\documentclass[a4paper,prb,superscriptaddress, amsfonts, amssymb, amsmath, reprint, showkeys, nofootinbib, twoside, twocolumn]{revtex4-2}
\usepackage[english]{babel}
\usepackage[utf8]{inputenc}
\usepackage{amsthm}
\usepackage{mathtools}
\usepackage{physics}
\usepackage{xcolor}
\usepackage{graphicx}
\usepackage[T1]{fontenc}
\usepackage[ pdftitle={Article}, pdfauthor={Author}]{hyperref} % For hyperlinks in the PDF
\usepackage{physics}
\usepackage[colorinlistoftodos, color=green!40, prependcaption]{todonotes}
\usepackage{dsfont}
\usepackage{chemformula}
\usepackage{units}
\usepackage{amsmath}
\usepackage{caption}
\usepackage{relsize}
\usepackage{enumitem}
\usepackage{epsfig}
\usepackage[amssymb]{SIunits}
\usepackage[normalem]{ulem}
\usepackage{esint}
\usepackage{epstopdf}
\usepackage{multirow}
\usepackage{csquotes}
\usepackage{mathrsfs}

\newcommand{\llint}[1]{\mathop{\mathlarger{\mathlarger{#1}}}}
\begin{document}
%\title{Electron-magnon scattering in complex systems from first principles} 
%\title{Nonlocal correlation effects due to virtual spin-flip processes in metallic collinear magnets} 
\title{Electronic correlations arising from anti-Stoner spin excitations: an \textit{ab initio} study of itinerant ferro- and antiferromagnet} 
\author{Sebastian Paischer} \email{sebastian.paischer@jku.at} 
\affiliation{Institute for  Theoretical Physics, Johannes Kepler  University Linz, Altenberger  Stra{\ss}e 69, 4040 Linz} 
%\author{Nadine Buczek} 
%\affiliation{Department of Applied Natural Sciences, L\"ubeck  University of Applied Sciences, M\"onkhofer Weg 239, 23562 L\"ubeck,  Germany} 
%\author{Giovanni Vignale}
%\affiliation{The Institute for Functional Intelligent Materials (I-FIM), National University of Singapore, 4 Science Drive 2, Singapore 117544 }
\author{David Eilmsteiner}
\affiliation{Institute for  Theoretical Physics, Johannes Kepler  University Linz, Altenberger  Stra{\ss}e 69, 4040 Linz}
\author{Mikhail I. Katsnelson} 
\affiliation{Institute for Molecules and Materials, Radboud University, Heyendaalseweg 135, 6525AJ Nijmegen, The Netherlands} 
\author{Arthur Ernst}
\affiliation{Institute for Theoretical Physics, Johannes Kepler University Linz, Altenberger  Stra{\ss}e 69, 4040 Linz}
\affiliation{Max Planck Institute of Microstructure Physics, Weinberg 2, D-06120 Halle, Germany}
\author{Pawe\l{} A. Buczek}
\affiliation{Department of Engineering and Computer Sciences, Hamburg  University of Applied Sciences, Berliner Tor 7, 20099 Hamburg, Germany} 

\date{\today}

\begin{abstract}
%In our recent report \cite{Paischer2023}, we introduced an efficient numerical scheme for calculating the electron-magnon self-energy in itinerant electron ferromagnets. Building on this foundation, we have now generalized our approach to encompass other collinear magnetic materials. This advancement is made possible by incorporating anti-Stoner processes of the magnetic susceptibility which were previously neglected. This generalization not only enhances our understanding of (weak) ferromagnets but also enables the study of antiferromagnets. To illustrate the versatility of our method, we apply it to ferromagnetic fcc Nickel as well as the antiferromagnet CrSb.
The anti-Stoner excitations are a spin-flips in which, effectively, an electron is promoted from a minority to a majority spin state, i.e., complementary to Stoner excitations and spin-waves. Since their spectral power is negligible in strong itinerant ferromagnets and they are identically absent in the ferromagnetic Heisenberg model, their properties and role in correlating electrons were hardly investigated so far. On the other hand, they are present in weak ferromagnets, fcc Ni being a prominent example, and both types of spin-flips (down-to-up and up-to-down) must be treated on equal footing in systems with degenerate spin up and down bands, in particular antiferromagnets in which case we choose CrSb as a model system. For these two materials we evaluate the strength of the effective interaction between the quasiparticles and the gas of virtual spin-flip excitations. To this end, we compute the corresponding self-energy taking advantage of our novel efficient \textit{ab initio} numerical scheme \cite{Paischer2023}. We find that in Ni the band-structure renormalization due to the anti-Stoner processes is weaker than the one due to Stoner-type magnons in the majority spin channel but the two become comparable in the minority one. The effect can be traced back primarily to the spectral strength of the respective spin excitations and the densities of the final available quasiparticle states in the scattering process. For the antiferromagnet, the situation is more complex and we observe that the electron-magnon interaction is sensitive not only to these densities of states but critically to the spatial shapes of the coupling magnonic modes as well.
\end{abstract}

\keywords{}

\maketitle

\section{Introduction}
Antiferromagnets (AFMs) were predicted long ago as magnetic materials containing two magnetic sublattices with antiparallel orientation of their magnetic moments and zero net magnetization \cite{Neel1932}; now this term is frequently applied to any magnetically ordered state with zero net magnetization. Actually, there are much more AFMs than ferromagnets in nature, and the study of AFMs is a very important part of the modern physics of magnetism \cite{goodenough1963magnetism,vonsovsky1974magnetism,white1983quantum,spaldin2010magnetic}. The absence of stray fields and relatively weak sensitivity to external magnetic fields in comparison with ferro- and ferrimagnets can be a serious advantage for the emerging fields of spintronics and magnonics \cite{Baltz2018,Han2023}. Furthermore, their ultrafast spin dynamics grants spintronic device operation in the terahertz frequency range \cite{Rezende2019}. Among the recent experimentally realized milestones in spintronic and magnonic using AFMs are the detection of ultrafast generation of spin currents \cite{Qiu2020}, the coherent spin wave transport including optical excitation and detection \cite{Hortensius2021} and long distance magnon transport in layered van der Waals AFMs \cite{Xing2019}. Furthermore, several unexpected phenomena like anomalous Hall effects, unconventional transport and the emergence of skyrmions were recently observed \cite{Nakata2017,Bonbien2021,Smejkal2022}.

On the theoretical side, one of the most challenging problems especially for materials including itinerant electron magnetism lies in the correct description of the correlated electronic structure. The importance of correlations has been studied on a model basis \cite{kotliar1,halfmet,beyond} and experimentally through comparison with density functional theory (DFT) calculations \cite{Kim2018}. Generally, correlations essentially influence material properties on a microscopic and macroscopic level, as for certain transitions metals they are even responsible for their color \cite{Acharya2023}. Of particular interest is the coupling of electrons with spin-flip excitations which are known to be strong in itinerant electron ferromagnets \cite{Edwards1973,Hertz1973,halfmet,IK_AFM}. When it comes to AFMs it is conjectured that correlations play an essential role in the context of superconductivity in antiferromagnetic pnictides and cuprates 
\cite{cuprates1,cuprates2,cuprates3,pnictides}.

One of the main theoretical tools utilized to study correlated electrons is the dynamic mean field theory (DMFT) often used in combination with DFT \cite{Anisimov1997,ldaplusplus,Kotliar2}. Several studies of AFMs using DMFT can be found in the literature \cite{Lichtenstein2000,Sangiovanni2006,Miura2008}. The main drawback of the DMFT lies in the local approximation which causes a wave-vector independent self energy. One can go beyond DMFT using e.g. the dual fermion method considering the DMFT solution as zero-order approximation \cite{Rubtsov2008,Rubtsov2009,beyond}. Alternative approaches using many body perturbation theory (MBPT), often in the formulation of Hedin \cite{Hedin1965}, are able to account for non-locality, but are often only applied to very simple materials due to their numerical complexity \cite{Mueller2016,Nabok2021}. However, we recently proposed a method based on MBPT which is able to account for the non-locality of the self energy while still being numerically feasible for complex materials \cite{Paischer2023}. This was achieved by approximating complex quantities of MBPT with their counterparts in the theory of time dependent DFT \cite{Buczek2011}. While the formalism presented in \cite{Paischer2023} is only applicable to strong ferromagnets, it can easily be expanded to being able to account for all thinkable collinear magnetic structures as will be shown in this work. In principle, this generalization also allows to consider paramagnets, but this aspect goes beyond the scope of the current report.

This paper is organized as follows. In chapter II we derive in detail the generalization of our recent method to incorporate anti-Stoner contributions to the self energy. The influence of these anti-Stoner excitations on ferromagnetic Ni will be discussed in the first part of chapter III before the discussion of antiferromagnetic CrSb in the second part of chapter III. Chapter IV summarized the main points of this report.

\section{Theory}

\subsection{Electron-magnon scattering}

In our recent report \cite{Paischer2023} we presented an efficient method to account for the scattering of electron and magnons in electronic structure calculations. Formally, the methodology is based on the exact Hedin equations and the evaluation of the many-body self-energy arising due to the magnon-mediated electron-electron scattering. It was shown that in collinear magnets the electron-magnon interaction enters through the vertex function encompassing terms which include the repeated scattering of electrons and holes of opposite spin.

The self-energy, for both spin channels in a collinear magnet, is given by
\begin{align}
    \varSigma_{\sigma}(\mathit{1},\mathit{2})=\text{i}\mathcal{V}^{\sigma}_{e-m}(\mathit{1}^+,\mathit{2})G_{\Bar{\sigma}}(\mathit{1},\mathit{2})
\end{align}
where $\sigma\in\{\uparrow,\downarrow\}$ denotes the spin of electron and hole quaisparticles, $\Bar{\sigma}$ is the opposite spin of $\sigma$, $\mathcal{V}^{\sigma}_{e-m}$ denotes the effective electron-electron interaction mediated by magnons and $G$ is the many-body single particle Green's function. With the self-energy at hand, the interacting Green's function including the electron-magnon interaction is found using the Dyson equation
\begin{align}
    G_{\sigma}(\mathit{1},\mathit{2})=G_{\sigma}^0(\mathit{1},\mathit{2})+G_{\sigma}^0(\mathit{1},\mathit{3})\varSigma_{\sigma}(\mathit{3},\mathit{4})G_{\sigma}(\mathit{4},\mathit{2})
\end{align}
where $G^0$ is the Green function of a reference system which in the context of this work is the LSDA Green function. In our numerical scheme the latter quantity is found directly using the Korringa–Kohn–Rostoker multiple-scattering approach \cite{Hoffmann2020}.

Now, the numerical evaluation of $\mathcal{V}^{\sigma}_{e-m}$ is far from being trivial. While in the case of strong ferromagnets it is given directly by the transverse magnetic susceptibility \cite{Paischer2023}, in general it must be obtained through a suitable analytic continuation of the spin-spin propagator. In the case of a periodic solid it is given by
\begin{align}
    &\mathcal{V}_{e-m}^{\uparrow}(\vb*{\varrho}_1,\vb*{\varrho}_2,\vb*{q},z)=2K^{+}_{\text{xc}}(\vb*{\varrho}_1)K^{+}_{\text{xc}}(\vb*{\varrho}_2)\nonumber\\
    &\left(\llint{\int}\frac{\dd{E}^\prime}{2\pi}\frac{\mathscr{C}^-(\vb*{\varrho}_1,\vb*{\varrho}_2,\vb*{q},E^\prime)}{z-E^\prime+\text{i}\epsilon}-\llint{\int}\frac{\dd{E}^\prime}{2\pi}\frac{\mathscr{C}^{+}(\vb*{\varrho}_2,\vb*{\varrho}_1,\vb*{q},E^\prime)}{z+E^\prime-\text{i}\epsilon}\right)\nonumber\\
    &\mathcal{V}_{e-m}^{\downarrow}(\vb*{\varrho}_1,\vb*{\varrho}_2,\vb*{q},z)=2K^{-}_{\text{xc}}(\vb*{\varrho}_1)K^{-}_{\text{xc}}(\vb*{\varrho}_2)\nonumber\\&\left(\llint{\int}\frac{\dd{E}^\prime}{2\pi}\frac{\mathscr{C}^{+}(\vb*{\varrho}_1,\vb*{\varrho}_2,\vb*{q},E^\prime)}{z-E^\prime+\text{i}\epsilon}-\llint{\int}\frac{\dd{E}^\prime}{2\pi}\frac{\mathscr{C}^{-}(\vb*{\varrho}_2,\vb*{\varrho}_1,\vb*{q},E^\prime)}{z+E^\prime-\text{i}\epsilon}\right)
\end{align}
where $\vb*{q}$ is the lattice (quasi)momentum, $\vb*{\varrho}$ represents the spatial coordinates within the unit cell, and $z$ is the complex energy. For the discussions in later parts of this work the cell coordinate $\vb*{\varrho}$ is split into the position of the particular atom site in the  cell (the site index $s$) and the relative coordinate $\vb*{r}$ within the Voronoi cell of this atom:
\begin{align}\label{eqn_slfe_sites}
	\varSigma_{\sigma}(\vb*{\varrho},\vb*{\varrho}',\vb*{q},z)=\varSigma^{ss'}_{\sigma}(\vb*{r},\vb*{r}',\vb*{q},z).
\end{align}

The retarded transverse magnetic susceptibility $\chi^{\pm}$, yielding the magnon spectrum of the system, can be expressed in terms of the correlation functions $\mathscr{C}^{\pm}$ 
\begin{align}
	\chi^{\pm}\left( \vb*{\varrho}, \vb*{\varrho}^{\prime},\vb*{q}, z\right)=&\int \frac{d E^{\prime}}{2 \pi} \frac{\mathscr{C}^{\pm}\left(\vb*{\varrho}, \vb*{\varrho}^{\prime},\vb*{q}, E^{\prime}\right)}{z-E^{\prime}+\mathrm{i} 0^{+}}\nonumber\\
	&-\int \frac{d E^{\prime}}{2 \pi} \frac{\mathscr{C}^{\mp}\left( \vb*{\varrho}^{\prime}, \vb*{\varrho},\vb*{q}, E^{\prime}\right)}{z+E^{\prime}-\mathrm{i} 0^{+}}
\end{align}
The reverse relation in the low temperature regime is given by the fluctuation-dissipation theorem
\begin{align}\label{eqn_fluc_diss}
	\mathscr{C}^\pm(\vb*{\varrho},\vb*{\varrho}',\vb*{q},E)=\text{i}\Theta(E)&\left[\chi^\pm(\vb*{\varrho},\vb*{\varrho}',\vb*{q},E+\text{i}\varepsilon)\right.\nonumber\\
	&\left.-\chi^\pm(\vb*{\varrho}'
	,\vb*{\varrho},\vb*{q},E+\text{i}\varepsilon)^\star\right].
\end{align} 

The relation between $\mathscr{C}^\pm$ and $\chi^{\pm}$ is the foundation of the numerical expediency of our scheme as $\chi^{\pm}$ (itself being an intricate non-trivial many-body quantity) can be obtained within the more affordable linear-response time-dependent density functional theory \cite{Gross1990}. 

Using the spectral representation of $\mathcal{V}_{e-m}$ and performing the necessary complex contour integration, the self-energy can be calculated as:
\begin{align}
	\varSigma_{\sigma}(\vb*{\varrho},\vb*{\varrho}',\vb*{q},z)=\frac{\text{i}}{2\pi\Omega_{\text{BZ}}}\left(\varSigma^{\text{S}}_{\sigma}(\vb*{\varrho},\vb*{\varrho}',\vb*{q},z)+\varSigma^{\text{aS}}_{\sigma}(\vb*{\varrho},\vb*{\varrho}',\vb*{q},z)\right)
\end{align}
where the Stoner (S) and anti-Stoner (aS) parts of the magnon mediated self-energy are given by
\begin{align}\label{eqn_slfe_S_aS}
	\varSigma_{\uparrow}^{\text{S}}&(\vb*{\varrho},\vb*{\varrho}',\vb*{q},z)=\int\dd[3]{q'}\sqint_{E<E_{\mathrm{F}}}\dd{z}'G_{\downarrow}(\vb*{\varrho},\vb*{\varrho}',\vb*{q}',z')\nonumber\\
	&\times\int\frac{\dd{E'}}{2\pi}\frac{2K_{\text{xc}}^{+}(\vb*{r})\mathscr{C}^+(\vb*{\varrho},\vb*{\varrho}',\vb*{q}-\vb*{q}',E')K_{\text{xc}}^{+}(\vb*{\varrho}')}{z-z'+E'-\text{i}\varepsilon}\nonumber\\
	\varSigma_{\uparrow}^{\text{aS}}&(\vb*{\varrho},\vb*{\varrho}',\vb*{q},z)=\int\dd[3]{q'}\sqint_{E>E_{\mathrm{F}}}\dd{z}'G_{\downarrow}(\vb*{\varrho},\vb*{\varrho}',\vb*{q}',z')\nonumber\\
	&\times\int\frac{\dd{E'}}{2\pi}\frac{2K_{\text{xc}}^+(\vb*{\varrho}')\mathscr{C}^-(\vb*{\varrho}',\vb*{\varrho},\vb*{q}-\vb*{q}',E')K_{\text{xc}}^+(\vb*{\varrho})}{z-z'-E'+\text{i}\varepsilon}\\
    \varSigma_{\downarrow}^{\text{S}}&(\vb*{\varrho},\vb*{\varrho}',\vb*{q},z)=\int\dd[3]{q'}\sqint_{E>E_{\mathrm{F}}}\dd{z}'G_{\uparrow}(\vb*{\varrho},\vb*{\varrho}',\vb*{q}',z')\nonumber\\
	&\times\int\frac{\dd{E'}}{2\pi}\frac{2K_{\text{xc}}^{-}(\vb*{r})\mathscr{C}^+(\vb*{\varrho},\vb*{\varrho}',\vb*{q}-\vb*{q}',E')K_{\text{xc}}^{-}(\vb*{\varrho}')}{z-z'-E'+\text{i}\varepsilon}\nonumber\\
	\varSigma_{\downarrow}^{\text{aS}}&(\vb*{\varrho},\vb*{\varrho}',\vb*{q},z)=\int\dd[3]{q'}\sqint_{E<E_{\mathrm{F}}}\dd{z}'G_{\uparrow}(\vb*{\varrho},\vb*{\varrho}',\vb*{q}',z')\nonumber\\
	&\times\int\frac{\dd{E'}}{2\pi}\frac{2K_{\text{xc}}^-(\vb*{\varrho}')\mathscr{C}^-(\vb*{\varrho}',\vb*{\varrho},\vb*{q}-\vb*{q}',E')K_{\text{xc}}^-(\vb*{\varrho})}{z-z'+E'-\text{i}\varepsilon}\nonumber
\end{align}
In the latter equation, the momentum ($\vb*{q}'$) is integrated over the Brillouin zone while the energy is integrated along a box contour with all Kohn-Sham states above or below the Fermi energy enclosed within it \cite{Paischer2023}.

\begin{figure}
	\centering
	\includegraphics[width=0.45\textwidth]{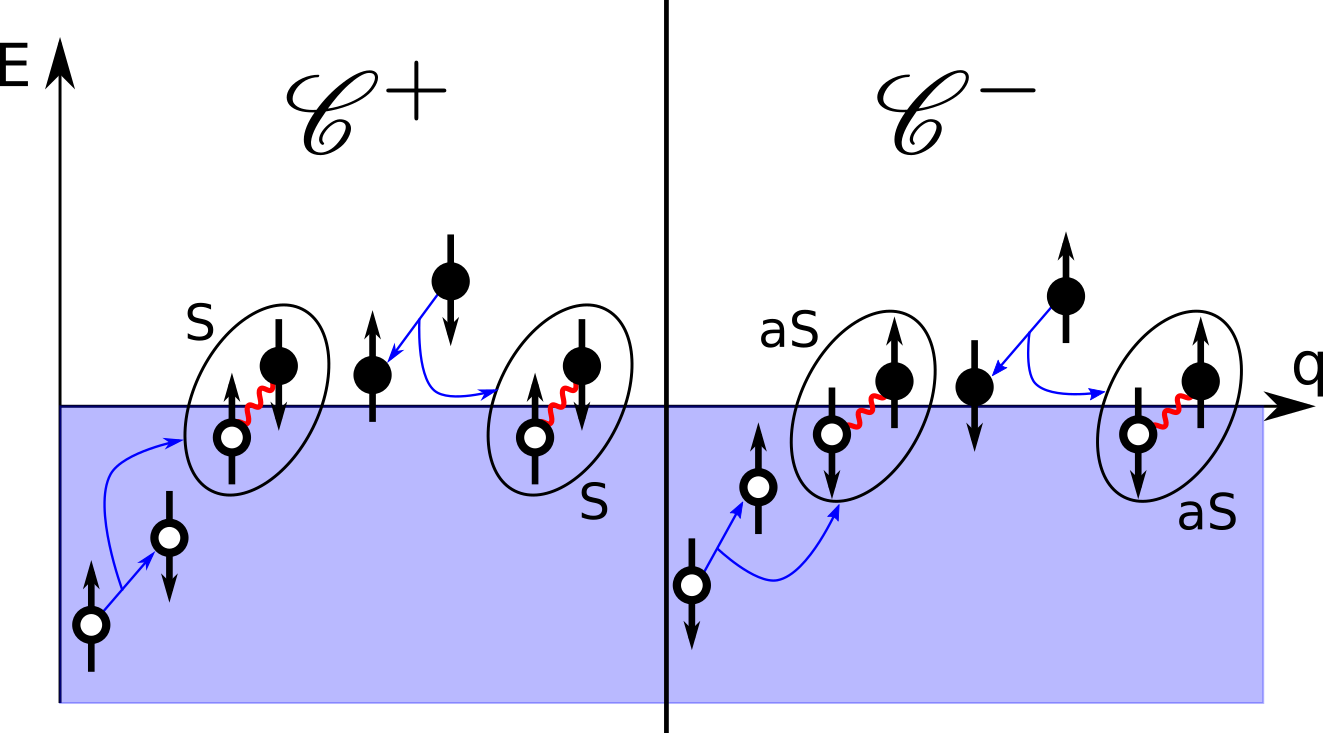}
	\caption{Schematics of possible quasiparticle processes caused by the Stoner (S) and anti-Stoner (aS) parts of the self-energy. At low temperatures all the processes depicted here are of Stokes character, i.e., they lead to the excitation of the electron liquid by a decaying quasiparticle. The two processes caused by $\mathscr{C}^+$ include a Stoner-like excitation while the two processes caused by $\mathscr{C}^-$ lead to the generation of an anti-Stoner pair.}
	\label{fig_processes}
\end{figure}

In the following, the physical interpretation of the magnon-electron processes contained in the above equations and the reason for introducing the Stoner (S) and anti-Stoner (aS) distinction in itinerant magnets is given. $\chi^+$, being the response to the anti-clockwise-polarized magnetic field oriented in the plane perpendicular to the ground state magnetization, describes, for positive frequencies, the up-to-down spin-flips. The spectral power of these processes is given by $\mathscr{C}^+$. They may be of single-particle character and, in ferromagnets with the ground state magnetization pointing up, are traditionally named \textit{Stoner excitations}. Typically, they feature a collective behavior known as \textit{spin-waves}. Since both of them are associated with a net flip of a single up-to-down electron spin, we refer to them as having \textit{Stoner character}. The possible electronic transitions due to the magnon mediated interaction caused by Stoner excitations are presented in the left part of Figure \ref{fig_processes}. Note, that at low temperatures considered solely in this work, all included process are of Stokes character, i.e., they involve an excitation of an electron-hole pair (or a magnon). At low temperatures, there are no magnetic excitations which could be absorbed by a quasiparticle, thus, there are no anti-Stokes processes.

$\mathscr{C}^-$ is the spectral power of down-to-up spin-flips, i.e., the processes of \textit{anti-Stoner character}. It is given by $\chi^-$ describing for positive frequencies the response to the clockwise-polarized magnetic field. The chirality of ferromagnets \cite{Sandratskii2012} yields $\mathscr{C}^-$ in these systems to be typically much smaller than $\mathscr{C}^+$. When $\mathscr{C}^-$ can be neglected whatsoever, one speaks of strong ferromagnets. The latter approximation is reasonable for insulating systems but not straightforwardly justifiable in the case of itinerant magnets. Obviously, it qualitatively breaks down for antiferromagnets and paramagnets with degenerate up and down bands. In the right part of figure \ref{fig_processes}, the two possible electronic transitions due to anti-Stoner excitations are depicted schematically.

The impact of electron-magnon scattering on the electronic structure of ferromagnets has been studied on the model level \cite{Edwards1973,Hertz1973,Vignale1985b,Ng1986,halfmet} and, more recently, in an ab initio and many-body context \cite{Mueller2019,Nabok2021,Paischer2023}. In all these studies, the impact of $\mathscr{C}^-$ was hardly addressed. The $\mathscr{C}^-=0$ approximation simplifies calculations since in this case, as carefully discussed in our previous report, the effective interaction $\mathcal{V}_{e-m}$ can be expressed by the retarded $\chi^\pm$ susceptibility directly without the need of the correlation functions $\mathscr{C}^\pm$. However, as we show in this paper, the impact of anti-Stoner excitations in \enquote{weak} itinerant magnets, e.g. fcc Ni, is not negligible and leads to sizable renormalization of the band structure. Moreover, in the antiferromagnets, the impact of the up-to-down and down-to-up spin processes must be treated on the equal footing. Here, we choose the CrSb antiferromagnetic system as an example of ab initio studies of magnon-electron interactions for a specific material.

In summary, we generalize our recently proposed method \cite{Paischer2023} to also cover anti-Stoner contribution to the self-energy. This is achieved by decomposing the susceptibility (and, consequently, the effective magnon meditated electron-electron interaction as well) into the correlation functions $\mathscr{C}^\pm$. In turn, these correlation functions allow to calculate the Stoner and anti-Stoner parts of the self-energy according to equation \ref{eqn_slfe_S_aS} and are themselves obtainable from the linear response time-depenedent DFT. 

\subsection{Linear response time-dependent density functional theory}

Linear response TDDFT allows to compute the retarded counterpart of the reducible density-density response function which can be used to extract the $\mathscr{C}^{\pm}$ spectral functions. First, the formally non-interacting Kohn-Sham response function is computed. It is given by
\begin{align}
	\chi_{\text{KS}}^{ij}(\vb*{x}_1,\vb*{x}_2,E)=&\sum_{mn}\sigma^i_{\alpha\beta}\sigma^j_{\gamma\delta}(f_m-f_n)\nonumber\\
	&\frac{\phi_{m\alpha}(\vb*{x}_1)^\star \phi_{n\beta}(\vb*{x}_1)\phi_{n\gamma}(\vb*{x}_2)^\star \phi_{m\delta}(\vb*{x}_2)}{E-(E_{m\alpha}-E_{n\beta})+i\epsilon}
\end{align}
where the sum includes Kohn-Sham states with energy $E_{m\alpha}$ and state vector $\phi_{m\alpha}$ while $f_m$ represents the Fermi-Dirac statistic and $\sigma^i$ is the $i$th Pauli matrix. Greek indices represent spin directions. For the spin-flip channels addressed in this work, the transverse response is obtained by setting $i,j\in\{x,y\}$.\\
Second, the evaluation of the true (interacting) susceptibility can be done upon solving the susceptibility Dyson equation 
\begin{align}\label{eqn_susc_dyson}
	\chi^{ij}&(\vb*{x}_1,\vb*{x}_2,E)=\chi_{\text{KS}}^{ij}(\vb*{x}_1,\vb*{x}_2,E)+\sum_{m,n}\iint\dd[3]{x_3}\dd[3]{x_4}\nonumber\\
	&\chi_{\text{KS}}^{im}(\vb*{x}_1,\vb*{x}_3,E)K_{\text{xc}}^{mn}(\vb*{x}_3,\vb*{x}_4,E)\chi^{nj}(\vb*{x}_4,\vb*{x}_2,E).
\end{align}
taking into account the changes of charge and magnetization densities or, equivalently, the Hartree and exchange-correlation potentials \cite{Buczek2011}. In the case of the adiabatic local spin density approximation \cite{Buczek2020} to the exchange-correlation kernel, the transverse magnetization response does not involve the charge or the Hartree response.

\subsection{Magnon-pole approximation and double counting}

%ICH GLAUBE DIESER ABSCHNITT WIRD NUR VERWIRREND FÜR DIE LESER SEIN. Wir nutzen die Näherung für C- nicht und für C+ ist sie schon woanders diskutiert. Ich würde den Abschnitt rausnehmen und Elemente davon an andern Stellen einbauen.

As discussed in our earlier work \cite{Paischer2023}, we resort to the \enquote{magnon-pole approximation} (unless otherwise specified) which replaces the TDDFT susceptibility with the susceptibility of the Heisenberg ferromagnet with exchange parameters calculated from the magnetic force theorem \cite{Lichtenstein1987}. There are two justifications for this approach. First, it was shown that through the solution of the susceptibility Dyson equation \ref{eqn_susc_dyson} the majority of the spectral weight of the Stoner excitations is shifted towards the spin-waves \cite{Buczek2011}. Second, the usage of the Heisenberg susceptibility effectively eliminates the typical double-counting problem encountered in similar theories. The problem arises due to the inclusion of an additional diagram which suffers from double counting \cite{Paischer2023}. One can deal with this problem by subtraction of the problematic term which on the flipside can result in violations of causality \cite{Mueller2019}. However, it can be shown that this spurious diagram represents the non-interacting part of the susceptibility, or in the language of TDDFT the Kohn-Sham susceptibility. Hence, the susceptibility required for the calculation of the electron-magnon self energy includes spin-waves while neglecting the single particle Stoner excitations. Consequently, the magnon-pole  turns out to be natural and physically well-motivated approximation. Finally, we note that due to its numerical complexity, our scheme is a one-shot method.

\section{Results}

\subsection{Anti-Stoner processes in fcc nickel}

%An example of the correlation functions for a specific wave-vector $\vb*{q}=(0.5,0.25,-0.25)\frac{2\pi}{a}$ is shown in Fig. \ref{fig_Ni_C} together with its single-particle counterpart given by the (non-interacting) Kohn-Sham correlation function. Through inclusion of the interaction a low energy spin-wave peak appears in the Stoner part (or in other words in $\mathscr{C}^+$). 

\begin{figure}
    \centering
    \includegraphics[width=0.45\textwidth]{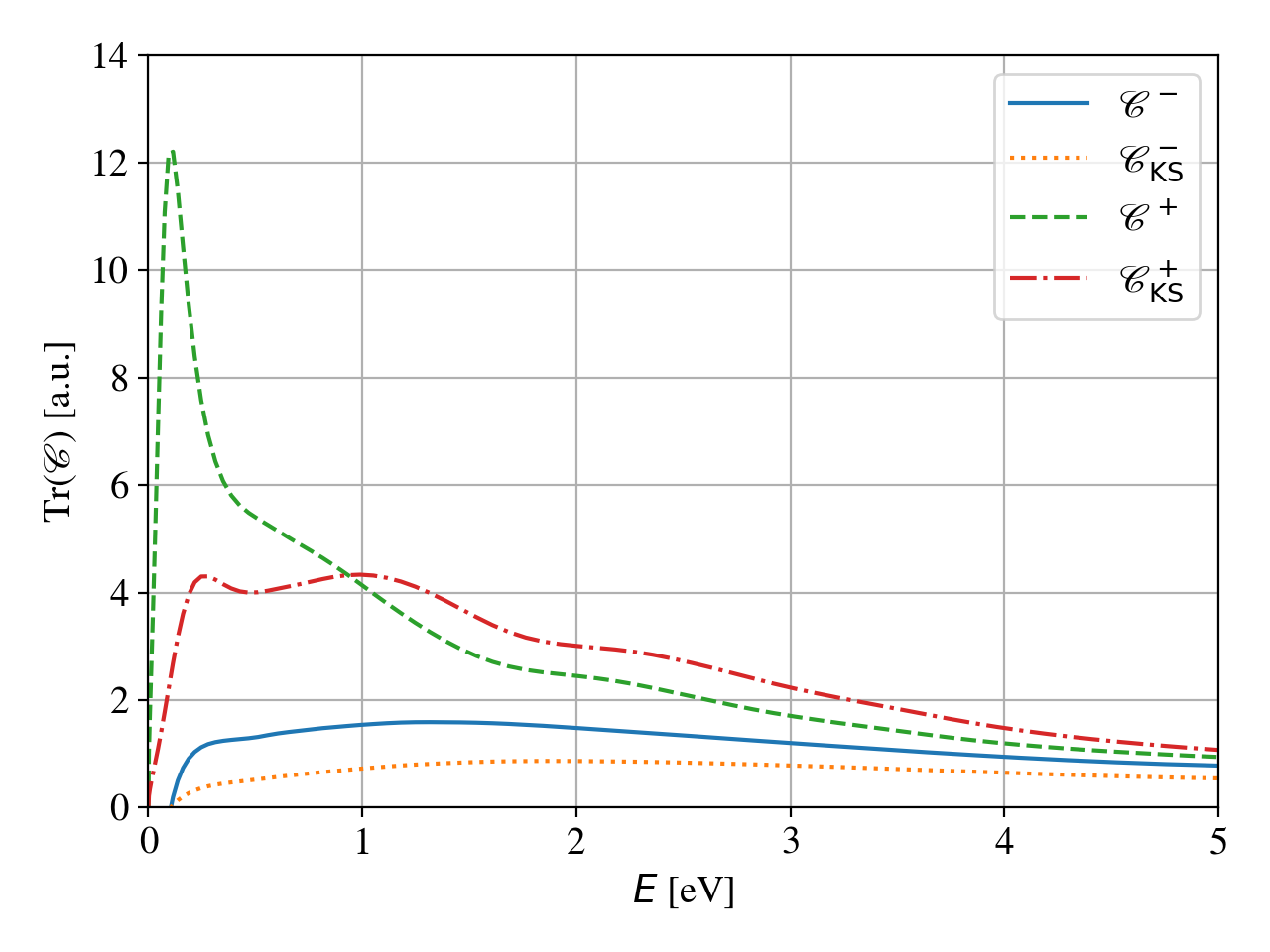}
    \caption{Spectral functions of the interacting and the Kohn-Sham (KS) system in fcc Nickel at $\vb*{q}=(0.5,0.25,-0.25)\frac{2\pi}{a}$.}
    \label{fig_Ni_C}
\end{figure}

Virtually no attention has been paid in the literature to the anti-Stoner excitations in itinerant ferromagnets.
%\emph{Thats technically true, but most works consider them implicitly. The difference is that within our formulism we can consider Stoner and anti-Stoner processes independently.}}
The signature of these excitations is found in the $\mathscr{C}^-$ spectral function and we recall that they correspond to effective down-to-up electron spin-flips of the Fermi liquid, cf. Fig. \ref{fig_processes}. For bcc Ni, an example of their spectrum is presented in the Fig. \ref{fig_Ni_C} for $\vb*{q}=(0.5,0.25,-0.25)\frac{2\pi}{a}$ alongside their Stoner counterpart.
The plots show that the anti-Stoner excitations ($\mathscr{C}^-$) are of primarily single-particle character in both Kohn-Sham and interacting system, i.e. no collective modes or sharp features are present. This is in striking contrast to the formation of spin-waves from the interacting Stoner excitations in its $\mathscr{C}^+$ counterpart. Furthermore, the spectral power contained in $\mathscr{C}^-$, as evident from Fig. \ref{fig_Ni_C}, is clearly smaller than in the Stoner case. However, as will be shown below, the impact of the quasiparticle scattering of these excitations leads to a sizable renormalization of the electronic band structure. Thus, the $\mathscr{C}^-$ contribution to the self-energy must be systematically incorporated in the description of itinerant ferromagnets.

We proceed now with the evaluation of the self-energy without the $\mathscr{C}^-$ = 0 approximation. 
%While the Stoner part of the self-energy is treated as a Heisenberg ferromagnet, 
due to the lack of well-defined collective modes in the $\mathscr{C}^-$-channel, no \enquote{magnon-pole}-like approximation can be constructed here, as it is done for the $\mathscr{C}^+$ spectral function \cite{Paischer2023}. Instead, $\mathscr{C}^-$ is taken directly from the linear response TDDFT. In order to account for the double counting issue \cite{Mueller2019}, the Kohn-Sham part of the correlation function must be subtracted from its enhanced counterpart, yielding
\begin{align}
\mathscr{C}^-=\mathscr{C}^-_{\mathrm{TDDFT}}-\mathscr{C}^-_{\mathrm{KS}}.
\end{align}
In general, this necessary correction may result in the wrong analytic structure of the self-energy (the \enquote{causality violation} \cite{Mueller2019}) but this is not the case for fcc Ni in the energy range considered. However, the situation is different for bcc iron (not studied here).

%The \emph{magnon-pole approximation} is based on the relation $\mathscr{C}^+\gg\mathscr{C}^-$ which holds for strong ferromagnets due to the higher density of majority spin carriers below the Fermi energy. Within our electron-magnon theory, this approximation shows itself most visibly in the imaginary part of the self energy which is zero for majority/minority carriers above/below the Fermi energy. 

\begin{figure}
	\centering
	\includegraphics[width=0.45\textwidth]{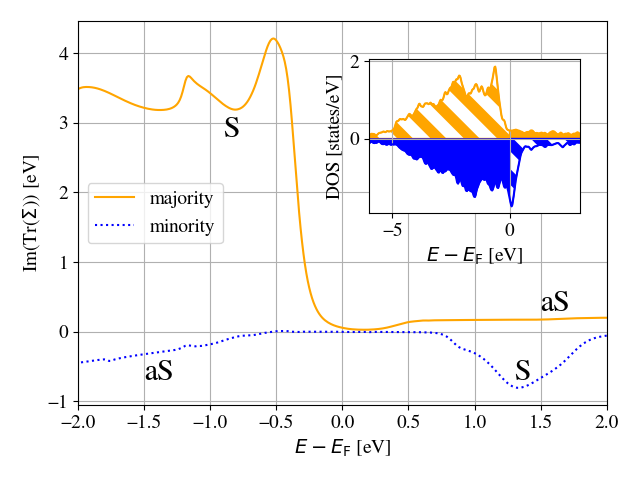}
	\caption{Trace of the imaginary part of the self-energy in fcc nickel at the $\Gamma$ point. The Stoner (S) contribution lead to finite quasiparticle life-time below/above the Fermi energy for the spin up/down channels. Anti-Stoner (aS) contributions cause the decay above/below the Fermi energy for spin up/down quasiparticles. The inset shows the (LDA) density of finite states in the decay processes involving the Stoner processes (fully filled) and anti-Stoner processes (hatched).}
    \label{fig_Ni_im_slfe}
\end{figure}

Upon the inclusion of anti-Stoner processes the imaginary part of the self-energy is non-zero both above and below the Fermi energy in both spin channels, cf. figure \ref{fig_Ni_im_slfe}. We recall that the $\mathscr{C}^- = 0$ (or \enquote{strong ferromagnet}) approximation precludes the magnon generation for the decaying spin-up electrons above the Fermi level and spin-down holes below it, cf. Fig. \ref{fig_processes}. Conversely, the decay of the majority electrons above the Fermi level and minority holes below the Fermi level is caused solely by the $\mathscr{C}^-$ processes.

The major impact of the electron-magnon scattering on the band structure occurs for majority holes below the Fermi level and is of Stoner character. This is in line with our previous report \cite{Paischer2023} and other studies \cite{SanchezBarriga2012,Mueller2019,Nabok2021} and can be explained by the spin asymmetry of the density of states, shown in the inset of figure \ref{fig_Ni_im_slfe}. There are both many final hole states (with spin down) and magnons to facilitate this intense scattering. The Stoner-type decay of minority electrons above the Fermi energy is impeded due to the scarcity of the unoccupied majority electronic states.

The anti-Stoner excitations can cause a decay of the majority spin carriers above the Fermi energy but the effect is small since there are both little final states for the decay and anti-Stoner excitations facilitating it.

When it comes to the minority carriers though, the self-energy is of similar magnitude above (governed by the Stoner processes) and below the Fermi energy (given by the anti-Stoner scattering), cf. figure \ref{fig_Ni_im_slfe}. This somewhat unexpected observation can be explained by considering the high density of the final majority holes which compensates for the reduced spectral intensity of $\mathscr{C}^-$.

%density of states, shown in the inset of figure \ref{fig_Ni_im_slfe}. For the minority self energy the final states of the electronic transition are majority spin states. Hence, the density of states above/below the Fermi energy is relevant for the Stoner/anti-Stoner contribution. On the other hand the correlation function $\mathscr{C}^+$/$\mathscr{C}^-$ gives rise to the Stoner/anti-Stoner part of the self energy. While the correlation function  for the Stoner part is clearly bigger than the one of the anti-Stoner part ($\mathscr{C}^+\gg\mathscr{C}^-$), the exact opposite is the case for the density of states of the final states. This leads to Stoner and anti-Stoner contributions of similar size for minority spin carriers. Note that the DOS of majority spin carriers corresponding to the Stoner part is larger than the one of then anti-Stoner part. Hence, the Stoner part of the self energy clearly dominates the majority self energy.

\begin{figure}
	\centering
	\includegraphics[width=0.45\textwidth]{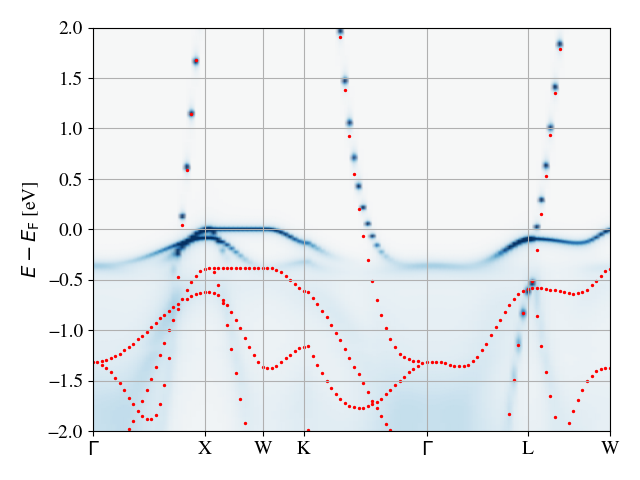}
	\includegraphics[width=0.45\textwidth]{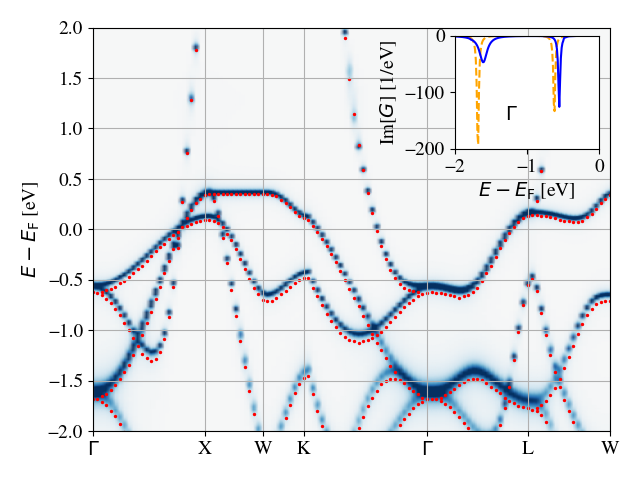}
	\caption{Electronic spectrum of fcc nickel within the LDA (red dots) and the LDA+$\mathcal{V}_{e-m}$ (blue background). Top: spin up. Bottom: spin down. Both Stoner and anti-Stoner processes are included. The broadening of the occupied down band is caused by the anti-Stoner scattering. The inset shows the imaginary part of the Green function within the LDA (dashed orange line) and the LDA+$\mathcal{V}_{e-m}$ (solid blue line) at the $\Gamma$ point.}
	\label{fig_Ni_bsf}
\end{figure}
When the electronic spectrum of nickel is considered, cf. Figure \ref{fig_Ni_bsf}, one observes that for spin up the spectrum remains mostly unchanged compared to earlier results \cite{Paischer2023}. On the other hand, the spectrum of spin-down quasiparticles becomes clearly renormalized. The energies of the bands are shifted by about 100\ {meV} and the hole states become damped with the strength growing with the distance to the Fermi level. We note that the result is consistent with the study based on the dynamical mean field theory \cite{SanchezBarriga2012}.

\subsection{Antiferromagnet CrSb}

The electron-magnon scattering in itinerant antiferromagnets has not been explored in the literature so far. At the model level, the effects of magnon-electron interaction on electronic properties of AFMs was considered in detail in Ref. \onlinecite{IK_AFM} where it was shown that magnons with wave vectors close to the antiferromagnetic wave vector play a crucial role. In this paper we investigate the phenomenon using our \textit{ab initio} methodology outlined above. Formally, there is no difference in the way the corresponding self-energy is evaluated for an AFM system compared to the case of the anti-Stoner contributions in an FM system as Ni above. However, an AFM, with its magnetically ordered both up and down degenerate sublattices, necessarily features equally intense up-to-down and down-to-up spin fluctuations.

We consider the layered AFM CrSb in the hexagonal NiAs structure. Due to its high N\'eel temperature this material has been identified as particularly promising in the field of AFM-spintronics. Furthermore, recent studies have shown that thin films of CrSb are ferrimagnetic or can even feature altermagnetic behavior \cite{Park2020,Reimers2024}. Here, we investigate the impact of electron-magnon coupling on the electronic properties of its bulk phase.

\begin{figure}[h]
	\centering
	\includegraphics[width=0.45\textwidth]{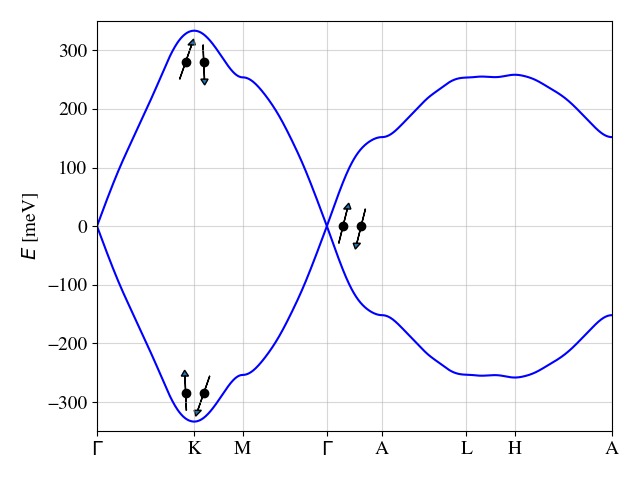}
    \includegraphics[width=0.45\textwidth]{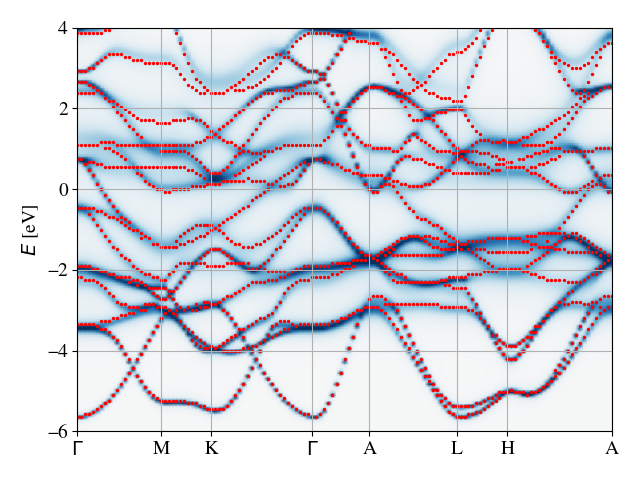}
	\caption{Upper panel: Magnonic spectrum of CrSb. The spatial shape of the magnon modes is shown schematically for the $\Gamma$ and the K points. Lower panel: LDA (red dots) and renormalized (blue background) electronc spectrum of CrSb.} 
	\label{fig_CrSb_magnons}
\end{figure}

The magnonic spectrum of CrSb is presented in Fig. \ref{fig_CrSb_magnons} (upper panel) with two spin-wave branches of linear dispersion close to the $\Gamma$-point featuring the Goldstone mode. These two branches are energetically degenerate but feature different spatial character indicated by the arrows close to the dispersion relations. The branch of positive frequency correspond to the effective up-to-down spin-flip while the negative frequency branch involves down-to-up spin-flip. In line with our discussion of the FM Ni system above, we could call them of Stoner and anti-Stoner part but one must stress that in an AFM, in contrast to the FM case, they are fully equivalent by symmetry. A further difference pertains to the fact that the collective spin-dynamics (magnons) emerges both in the Stoner and the anti-Stoner channel. %\sout{Our calculations show that the modes are only weakly Landau-damped and we resort to the magnon-pole approximation when evaluating the magnon-electron self-energy below.}

The chirality of magnon excitations in AFM has already been carefully discussed before \cite{Sandratskii2012}. The magnons break the symmetry between the sublattices with upwards ($\Uparrow$) and downwards ($\Downarrow$) magnetization. The up-to-down spin-flip is associated with the up sublattice precessing stronger (with larger amplitude) than the down one. Conversely, the down-to-up spin-flip causes the down sublattice to precess stronger than the up one. In both cases the magnon causes the system to feature a net magnetic moment which is absent in the ground state. These intricate symmetries are reflected in the properties of the self-energy discussed below.

\begin{figure}[h]
	\centering
    \includegraphics[width=0.45\textwidth]{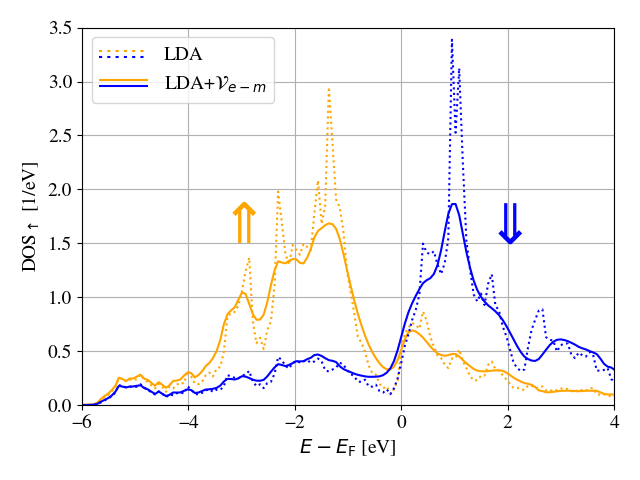}	
	\includegraphics[width=0.45\textwidth]{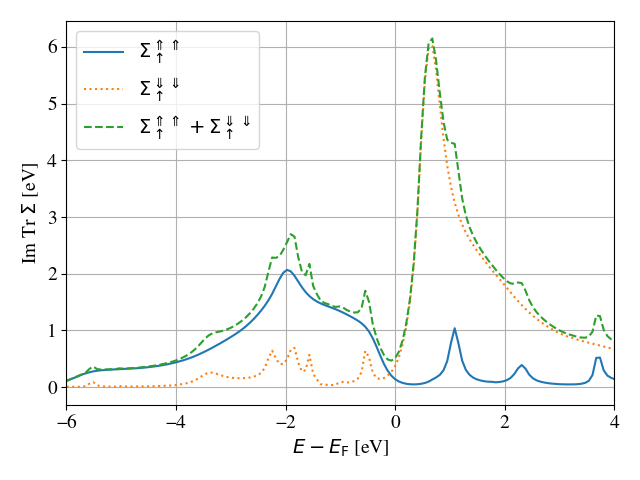}
	\caption{Upper panel: Density of states for up carriers at the two Cr atoms (upwards $\Uparrow$ and downwards $\Downarrow$ magnetization labeled in the figure) within the LDA and the LDA+$\mathcal{V}_{e-m}$. Lower panel: Trace of the self-energy (imaginary part) for up carriers at the $\Gamma$ point for the two magnetic Cr up and down sites.}
	\label{fig_CrSb_slfe}
\end{figure}

First, we note the symmetry relations between the up and down carrier self-energy:
\begin{align}
\varSigma_{\uparrow}^{\Uparrow\Uparrow}=&\left(\varSigma_{\downarrow}^{\Downarrow\Downarrow}\right)^\dagger\nonumber\\		\varSigma_{\uparrow}^{\Downarrow\Downarrow}=&\left(\varSigma_{\downarrow}^{\Uparrow\Uparrow}\right)^\dagger	
\end{align}
Here, the subscript labels the spin of the quasi-particle while the superscripts represent the site-resolved blocks of the self-energy, i.e., Cr site with upwards $\Uparrow$ or downwards $\Downarrow$ magnetization (see also equation \ref{eqn_slfe_sites}). These symmetries are caused by the fact that the electronic states for both spin directions are degenerate and the magnon spectrum obeys the symmetries discussed above. Consequently, in antiferromagnets the electronic spectra of both spin channels are renormalized in the same manner upon the action of the magnon-electron interaction. In the specific case of CrSb, the energy of electronic bands changes weakly, cf. lower panel in Fig. \ref{fig_CrSb_magnons}. However, they experience rather strong damping caused by the scattering with the virtual magnons. Note that the damping of the electronic states is strongly dependent on the momentum of the state.

%\begin{figure}[h]
%	\centering
%	%\includegraphics[width=0.45\textwidth]{pics/CrSb_bsf.png}
%	\includegraphics[width=0.45\textwidth]{pics/CrSb_dos.png}		
%	\caption{Density of states for up carriers at the two Cr atoms (upwards $\Uparrow$ and downwards $\Downarrow$ magnetization labeled in the figure) within the LDA and the LDA+$\mathcal{V}_{e-m}$.}
%	\label{fig_CrSb_dos}
%\end{figure}

\begin{figure}[h]
	\centering
	\includegraphics[width=0.45\textwidth]{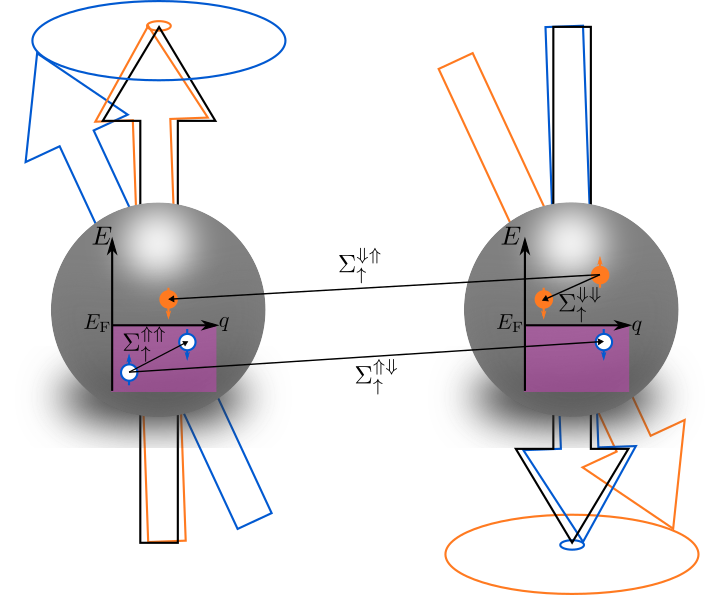}
	\caption{Schematic picture of the dominating electronic transitions in AFMs due to the electron-magnon interactions. The large black background arrows represent magnetic moments of the corresponding site. Two magnon modes with equal and finite momentum are shown schematically in blue and orange. These magnon modes are strongly located at one of the sites as can be seen by the tilt of the precessing magnetic moments. The electronic transitions due to magnons are shown in the color of the magnon which causes them. For example, the blue magnon allows an up hole to transit efficiently into a down hole state on the same site (leading to the large value of $\Im \varSigma_{\uparrow}^{\Uparrow\Uparrow}(E < E_{F})$). }
	\label{fig_afm}
\end{figure}

In what follows, we analyze more closely the properties of the quasiparticle decay channels opening due to their interaction with magnons in antiferromagnets.

First of all, contrary to the case of strong FM systems, both up and down electrons and holes can involve in a sizable interaction with magnons. A further difference is manifested by the fact that in the FM systems the strength of the scattering is primarily determined by the availability of the final quasiparticle states while for antiferromagnets the spatial shape of the magnon modes facilitating the decay is of paramount importance as well.

The site resolved density of states, cf. upper panel in figure \ref{fig_CrSb_slfe}, shows that the density of states of up spin particles is high/low below/above the Fermi level on the site with upwards magnetization while the opposite is true for the site with downwards magnetization, cf.\ also \cite{Dijkstra1989,Ito2007,Kahal2007,Polesya2011}.

The lower panel in Fig.\ \ref{fig_CrSb_slfe} shows an example of the CrSb self-energy (imaginary part leading to the finite quasiparticle life-time) for up ($\uparrow$) charge carriers. There is a clear asymmetry between electron and hole dynamics on different Cr sites (hosting up $\Uparrow$ and down $\Downarrow$ magnetic moments respectively). Up electrons on the Cr$_{\Uparrow}$ interact weakly with magnons, featuring a small imaginary part of the self-energy, and, consequently, long life-time. This effect can be traced back to spatial form of the magnon facilitating the decay. The magnon must be of anti-Stoner character and thus its amplitude on the Cr$_{\Uparrow}$ is small. This limits the decay despite the fact that there are many available unoccupied final electron down states on this site. Note that the maxima in the density of final states are roughly reflected in the moderate isolated peaks of $\Im \varSigma_{\uparrow}^{\Uparrow\Uparrow}(E > E_{F})$). An equivalent argumentation explains the long life-times of up holes on the Cr$_{\Downarrow}$ sites.

On the other hand, the up holes on the Cr$_{\Uparrow}$ site can emit magnons and decay efficiently, $\Im \varSigma_{\uparrow}^{\Uparrow\Uparrow}(E < E_{F})$ is large. Their decay process requires Stoner-like excitations of the Fermi liquid which feature large amplitudes on the Cr$_{\Uparrow}$ sites. While the density of final down holes is limited on the site, it is still sizable. The intensity of the decay processes of the up electrons on the Cr$_{\Downarrow}$ sites can be qualitatively explained using this same argument. Of course, the carriers can also decay into states on other Cr sites, the processes being described by the off-diagonal parts of the self-energy $\varSigma_{\sigma}^{\Uparrow\Downarrow}$ and $\varSigma_{\sigma}^{\Downarrow\Uparrow}$. These dominating quasiparticle-magnon scattering processes are depicted schematically in Fig.\ \ref{fig_afm}.

With the self-energy at hand, the renormalized Fermi level and the magnetic moments can be computed. Due to the high numerical cost, this is done in a one-shot procedure, i.e., not self-consistently. It turns out that the Fermi energy decreases by approximately 0.2\ eV. The magnetic moment of the Cr atoms is reduced from 2.84\ $\mu_{\text{B}}$ (LSDA approximation) to 2.67\ $\mu_{\text{B}}$ (LSDA+$\mathcal{V}_{e-m}$ approximation including the magnon-electron scattering) which agrees with available experimental results \cite{Snow1952}. The renormalized density of states is shown in the upper panel in Fig.\ \ref{fig_CrSb_slfe}.

We conclude by briefly discussing the role of the non-magnetic antimony atoms in the magnon-electron scattering. When resorting to the magnon-pole approximation, their contribution vanishes identically, as they carry no magnetic moment in the ground state. Furthermore, it means that the DOS of the Sb sites is unchanged in this approximation. However, in general the non-magnetic atoms could become magnetically polarized when neighboring magnetic moments cease to be collinear during a magnon excitation and the approximation might break. Nevertheless, the magnetic polarizability of Sb in the CrSb system is minuscule which we confirmed by computing the full magnetic susceptibility from linear response TDDFT for selected wave-vectors. On the other hand, all atoms, including the Sb, are fully included in the description of the electronic structure given by the KKR Green's function.

\section{Summary}
In this report, we completed the \textit{ab initio} description of magnon-electron scattering processes in collinear magnets. On the example of fcc Ni, we investigated carefully the correlations emerging from the interaction of quasiparticles and the gas of virtual anti-Stoner excitations. While its influence is weaker than the one of Stoner-type magnons, our results show that the strong ferromagnet approximation does not hold universally for magnon-induced electron damping in all itinerant ferromagnets. In the particular case of fcc Ni, the impact of Stoner- and anti-Stoner-like exciations on the quasiparticle spectra becomes of comaprable strength in the minority spin channel while in the majority one the Stoner-type excitations dominate.

In antiferromagnetic systems, as shown on the example of CrSb, the electron-magnon interaction is sensitive not only to the density of available final particle states in the scattering processes, but to the spatial shape of coupling magnonic modes as well. This result is in striking contrast to ferromagnetic materials like nickel for which the density of final states is the main factor impacting the strength of the quasiparticle renormalization. Furthermore, the antiferromagnetic symmetries of the electronic structure and chirality of the magnon excitations lead to the quasiparticle damping which acts selectively based on the atomic site and its magnetic moment. We believe it to offer a new approach to carrier life-time engineering in spintronic devices.
 
\section*{Acknowledgments}
S.P. is recipient of a DOC Fellowship of the Austrian Academy of Sciences at the Institute of mathematics, physics, space research and materials sciences. A.E. acknowledges the funding by the Fonds zur F\"orderung der Wissenschaftlichen Forschung (FWF) under Grant No. I 5384. P.B. gratefully acknowledge financial support from the DFG-LAV grant \enquote{SPINELS} and HSP grant \enquote{DEUM}. The work of M.I.K. was supported by the European Union’s Horizon 2020 research and innovation program under European Research Council synergy grant 854843 \enquote{FASTCORR}.

\bibliographystyle{apsrev4-2}
\bibliography{./Quellen}

\begin{thebibliography}{55}%
\makeatletter
\providecommand \@ifxundefined [1]{%
 \@ifx{#1\undefined}
}%
\providecommand \@ifnum [1]{%
 \ifnum #1\expandafter \@firstoftwo
 \else \expandafter \@secondoftwo
 \fi
}%
\providecommand \@ifx [1]{%
 \ifx #1\expandafter \@firstoftwo
 \else \expandafter \@secondoftwo
 \fi
}%
\providecommand \natexlab [1]{#1}%
\providecommand \enquote  [1]{``#1''}%
\providecommand \bibnamefont  [1]{#1}%
\providecommand \bibfnamefont [1]{#1}%
\providecommand \citenamefont [1]{#1}%
\providecommand \href@noop [0]{\@secondoftwo}%
\providecommand \href [0]{\begingroup \@sanitize@url \@href}%
\providecommand \@href[1]{\@@startlink{#1}\@@href}%
\providecommand \@@href[1]{\endgroup#1\@@endlink}%
\providecommand \@sanitize@url [0]{\catcode `\\12\catcode `\$12\catcode
  `\&12\catcode `\#12\catcode `\^12\catcode `\_12\catcode `\%12\relax}%
\providecommand \@@startlink[1]{}%
\providecommand \@@endlink[0]{}%
\providecommand \url  [0]{\begingroup\@sanitize@url \@url }%
\providecommand \@url [1]{\endgroup\@href {#1}{\urlprefix }}%
\providecommand \urlprefix  [0]{URL }%
\providecommand \Eprint [0]{\href }%
\providecommand \doibase [0]{https://doi.org/}%
\providecommand \selectlanguage [0]{\@gobble}%
\providecommand \bibinfo  [0]{\@secondoftwo}%
\providecommand \bibfield  [0]{\@secondoftwo}%
\providecommand \translation [1]{[#1]}%
\providecommand \BibitemOpen [0]{}%
\providecommand \bibitemStop [0]{}%
\providecommand \bibitemNoStop [0]{.\EOS\space}%
\providecommand \EOS [0]{\spacefactor3000\relax}%
\providecommand \BibitemShut  [1]{\csname bibitem#1\endcsname}%
\let\auto@bib@innerbib\@empty
%</preamble>
\bibitem [{\citenamefont {Paischer}\ \emph {et~al.}(2023)\citenamefont
  {Paischer}, \citenamefont {Vignale}, \citenamefont {Katsnelson},
  \citenamefont {Ernst},\ and\ \citenamefont {Buczek}}]{Paischer2023}%
  \BibitemOpen
  \bibfield  {author} {\bibinfo {author} {\bibfnamefont {S.}~\bibnamefont
  {Paischer}}, \bibinfo {author} {\bibfnamefont {G.}~\bibnamefont {Vignale}},
  \bibinfo {author} {\bibfnamefont {M.~I.}\ \bibnamefont {Katsnelson}},
  \bibinfo {author} {\bibfnamefont {A.}~\bibnamefont {Ernst}},\ and\ \bibinfo
  {author} {\bibfnamefont {P.~A.}\ \bibnamefont {Buczek}},\ }\bibfield
  {journal} {\bibinfo  {journal} {Physical Review B}\ }\textbf {\bibinfo
  {volume} {107}},\ \href {https://doi.org/10.1103/physrevb.107.134410}
  {10.1103/physrevb.107.134410} (\bibinfo {year} {2023})\BibitemShut {NoStop}%
\bibitem [{\citenamefont {Néel}(1932)}]{Neel1932}%
  \BibitemOpen
  \bibfield  {author} {\bibinfo {author} {\bibfnamefont {L.}~\bibnamefont
  {Néel}},\ }\href {https://doi.org/10.1051/anphys/193210180005} {\bibfield
  {journal} {\bibinfo  {journal} {Annales de Physique}\ }\textbf {\bibinfo
  {volume} {10}},\ \bibinfo {pages} {5} (\bibinfo {year} {1932})}\BibitemShut
  {NoStop}%
\bibitem [{\citenamefont {Goodenough}(1963)}]{goodenough1963magnetism}%
  \BibitemOpen
  \bibfield  {author} {\bibinfo {author} {\bibfnamefont {J.~B.}\ \bibnamefont
  {Goodenough}},\ }\href@noop {} {\emph {\bibinfo {title} {{Magnetism and the
  Chemical Bond}}}}\ (\bibinfo  {publisher} {New York: Interscience
  Publishers},\ \bibinfo {year} {1963})\BibitemShut {NoStop}%
\bibitem [{\citenamefont {Vonsovskii}(1974)}]{vonsovsky1974magnetism}%
  \BibitemOpen
  \bibfield  {author} {\bibinfo {author} {\bibfnamefont {S.~V.}\ \bibnamefont
  {Vonsovskii}},\ }\href@noop {} {\emph {\bibinfo {title} {{Magnetism}}}},\
  Vol.~\bibinfo {volume} {2}\ (\bibinfo  {publisher} {New York: J. Wiley \&
  Sons},\ \bibinfo {year} {1974})\BibitemShut {NoStop}%
\bibitem [{\citenamefont {White}\ and\ \citenamefont
  {Bayne}(1983)}]{white1983quantum}%
  \BibitemOpen
  \bibfield  {author} {\bibinfo {author} {\bibfnamefont {R.~M.}\ \bibnamefont
  {White}}\ and\ \bibinfo {author} {\bibfnamefont {B.}~\bibnamefont {Bayne}},\
  }\href@noop {} {\emph {\bibinfo {title} {{Quantum theory of magnetism}}}},\
  Vol.~\bibinfo {volume} {1}\ (\bibinfo  {publisher} {Springer},\ \bibinfo
  {year} {1983})\BibitemShut {NoStop}%
\bibitem [{\citenamefont {Spaldin}(2010)}]{spaldin2010magnetic}%
  \BibitemOpen
  \bibfield  {author} {\bibinfo {author} {\bibfnamefont {N.~A.}\ \bibnamefont
  {Spaldin}},\ }\href@noop {} {\emph {\bibinfo {title} {{Magnetic materials:
  fundamentals and applications}}}}\ (\bibinfo  {publisher} {Cambridge
  university press},\ \bibinfo {year} {2010})\BibitemShut {NoStop}%
\bibitem [{\citenamefont {Baltz}\ \emph {et~al.}(2018)\citenamefont {Baltz},
  \citenamefont {Manchon}, \citenamefont {Tsoi}, \citenamefont {Moriyama},
  \citenamefont {Ono},\ and\ \citenamefont {Tserkovnyak}}]{Baltz2018}%
  \BibitemOpen
  \bibfield  {author} {\bibinfo {author} {\bibfnamefont {V.}~\bibnamefont
  {Baltz}}, \bibinfo {author} {\bibfnamefont {A.}~\bibnamefont {Manchon}},
  \bibinfo {author} {\bibfnamefont {M.}~\bibnamefont {Tsoi}}, \bibinfo {author}
  {\bibfnamefont {T.}~\bibnamefont {Moriyama}}, \bibinfo {author}
  {\bibfnamefont {T.}~\bibnamefont {Ono}},\ and\ \bibinfo {author}
  {\bibfnamefont {Y.}~\bibnamefont {Tserkovnyak}},\ }\bibfield  {journal}
  {\bibinfo  {journal} {Reviews of Modern Physics}\ }\textbf {\bibinfo {volume}
  {90}},\ \href {https://doi.org/10.1103/revmodphys.90.015005}
  {10.1103/revmodphys.90.015005} (\bibinfo {year} {2018})\BibitemShut {NoStop}%
\bibitem [{\citenamefont {Han}\ \emph {et~al.}(2023)\citenamefont {Han},
  \citenamefont {Cheng}, \citenamefont {Liu}, \citenamefont {Ohno},\ and\
  \citenamefont {Fukami}}]{Han2023}%
  \BibitemOpen
  \bibfield  {author} {\bibinfo {author} {\bibfnamefont {J.}~\bibnamefont
  {Han}}, \bibinfo {author} {\bibfnamefont {R.}~\bibnamefont {Cheng}}, \bibinfo
  {author} {\bibfnamefont {L.}~\bibnamefont {Liu}}, \bibinfo {author}
  {\bibfnamefont {H.}~\bibnamefont {Ohno}},\ and\ \bibinfo {author}
  {\bibfnamefont {S.}~\bibnamefont {Fukami}},\ }\href
  {https://doi.org/10.1038/s41563-023-01492-6} {\bibfield  {journal} {\bibinfo
  {journal} {Nature Materials}\ }\textbf {\bibinfo {volume} {22}},\ \bibinfo
  {pages} {684} (\bibinfo {year} {2023})}\BibitemShut {NoStop}%
\bibitem [{\citenamefont {Rezende}\ \emph {et~al.}(2019)\citenamefont
  {Rezende}, \citenamefont {Azevedo},\ and\ \citenamefont
  {Rodríguez-Suárez}}]{Rezende2019}%
  \BibitemOpen
  \bibfield  {author} {\bibinfo {author} {\bibfnamefont {S.~M.}\ \bibnamefont
  {Rezende}}, \bibinfo {author} {\bibfnamefont {A.}~\bibnamefont {Azevedo}},\
  and\ \bibinfo {author} {\bibfnamefont {R.~L.}\ \bibnamefont
  {Rodríguez-Suárez}},\ }\bibfield  {journal} {\bibinfo  {journal} {Journal
  of Applied Physics}\ }\textbf {\bibinfo {volume} {126}},\ \href
  {https://doi.org/10.1063/1.5109132} {10.1063/1.5109132} (\bibinfo {year}
  {2019})\BibitemShut {NoStop}%
\bibitem [{\citenamefont {Qiu}\ \emph {et~al.}(2020)\citenamefont {Qiu},
  \citenamefont {Zhou}, \citenamefont {Zhang}, \citenamefont {Wu},
  \citenamefont {Tian}, \citenamefont {Cheng}, \citenamefont {Mi},
  \citenamefont {Zhao}, \citenamefont {Zhang}, \citenamefont {Wu},
  \citenamefont {Jin}, \citenamefont {Chen},\ and\ \citenamefont
  {Wu}}]{Qiu2020}%
  \BibitemOpen
  \bibfield  {author} {\bibinfo {author} {\bibfnamefont {H.}~\bibnamefont
  {Qiu}}, \bibinfo {author} {\bibfnamefont {L.}~\bibnamefont {Zhou}}, \bibinfo
  {author} {\bibfnamefont {C.}~\bibnamefont {Zhang}}, \bibinfo {author}
  {\bibfnamefont {J.}~\bibnamefont {Wu}}, \bibinfo {author} {\bibfnamefont
  {Y.}~\bibnamefont {Tian}}, \bibinfo {author} {\bibfnamefont {S.}~\bibnamefont
  {Cheng}}, \bibinfo {author} {\bibfnamefont {S.}~\bibnamefont {Mi}}, \bibinfo
  {author} {\bibfnamefont {H.}~\bibnamefont {Zhao}}, \bibinfo {author}
  {\bibfnamefont {Q.}~\bibnamefont {Zhang}}, \bibinfo {author} {\bibfnamefont
  {D.}~\bibnamefont {Wu}}, \bibinfo {author} {\bibfnamefont {B.}~\bibnamefont
  {Jin}}, \bibinfo {author} {\bibfnamefont {J.}~\bibnamefont {Chen}},\ and\
  \bibinfo {author} {\bibfnamefont {P.}~\bibnamefont {Wu}},\ }\href
  {https://doi.org/10.1038/s41567-020-01061-7} {\bibfield  {journal} {\bibinfo
  {journal} {Nature Physics}\ }\textbf {\bibinfo {volume} {17}},\ \bibinfo
  {pages} {388} (\bibinfo {year} {2020})}\BibitemShut {NoStop}%
\bibitem [{\citenamefont {Hortensius}\ \emph {et~al.}(2021)\citenamefont
  {Hortensius}, \citenamefont {Afanasiev}, \citenamefont {Matthiesen},
  \citenamefont {Leenders}, \citenamefont {Citro}, \citenamefont {Kimel},
  \citenamefont {Mikhaylovskiy}, \citenamefont {Ivanov},\ and\ \citenamefont
  {Caviglia}}]{Hortensius2021}%
  \BibitemOpen
  \bibfield  {author} {\bibinfo {author} {\bibfnamefont {J.~R.}\ \bibnamefont
  {Hortensius}}, \bibinfo {author} {\bibfnamefont {D.}~\bibnamefont
  {Afanasiev}}, \bibinfo {author} {\bibfnamefont {M.}~\bibnamefont
  {Matthiesen}}, \bibinfo {author} {\bibfnamefont {R.}~\bibnamefont
  {Leenders}}, \bibinfo {author} {\bibfnamefont {R.}~\bibnamefont {Citro}},
  \bibinfo {author} {\bibfnamefont {A.~V.}\ \bibnamefont {Kimel}}, \bibinfo
  {author} {\bibfnamefont {R.~V.}\ \bibnamefont {Mikhaylovskiy}}, \bibinfo
  {author} {\bibfnamefont {B.~A.}\ \bibnamefont {Ivanov}},\ and\ \bibinfo
  {author} {\bibfnamefont {A.~D.}\ \bibnamefont {Caviglia}},\ }\href
  {https://doi.org/10.1038/s41567-021-01290-4} {\bibfield  {journal} {\bibinfo
  {journal} {Nature Physics}\ }\textbf {\bibinfo {volume} {17}},\ \bibinfo
  {pages} {1001} (\bibinfo {year} {2021})}\BibitemShut {NoStop}%
\bibitem [{\citenamefont {Xing}\ \emph {et~al.}(2019)\citenamefont {Xing},
  \citenamefont {Qiu}, \citenamefont {Wang}, \citenamefont {Yao}, \citenamefont
  {Ma}, \citenamefont {Cai}, \citenamefont {Jia}, \citenamefont {Xie},\ and\
  \citenamefont {Han}}]{Xing2019}%
  \BibitemOpen
  \bibfield  {author} {\bibinfo {author} {\bibfnamefont {W.}~\bibnamefont
  {Xing}}, \bibinfo {author} {\bibfnamefont {L.}~\bibnamefont {Qiu}}, \bibinfo
  {author} {\bibfnamefont {X.}~\bibnamefont {Wang}}, \bibinfo {author}
  {\bibfnamefont {Y.}~\bibnamefont {Yao}}, \bibinfo {author} {\bibfnamefont
  {Y.}~\bibnamefont {Ma}}, \bibinfo {author} {\bibfnamefont {R.}~\bibnamefont
  {Cai}}, \bibinfo {author} {\bibfnamefont {S.}~\bibnamefont {Jia}}, \bibinfo
  {author} {\bibfnamefont {X.}~\bibnamefont {Xie}},\ and\ \bibinfo {author}
  {\bibfnamefont {W.}~\bibnamefont {Han}},\ }\bibfield  {journal} {\bibinfo
  {journal} {Physical Review X}\ }\textbf {\bibinfo {volume} {9}},\ \href
  {https://doi.org/10.1103/physrevx.9.011026} {10.1103/physrevx.9.011026}
  (\bibinfo {year} {2019})\BibitemShut {NoStop}%
\bibitem [{\citenamefont {Nakata}\ \emph {et~al.}(2017)\citenamefont {Nakata},
  \citenamefont {Kim}, \citenamefont {Klinovaja},\ and\ \citenamefont
  {Loss}}]{Nakata2017}%
  \BibitemOpen
  \bibfield  {author} {\bibinfo {author} {\bibfnamefont {K.}~\bibnamefont
  {Nakata}}, \bibinfo {author} {\bibfnamefont {S.~K.}\ \bibnamefont {Kim}},
  \bibinfo {author} {\bibfnamefont {J.}~\bibnamefont {Klinovaja}},\ and\
  \bibinfo {author} {\bibfnamefont {D.}~\bibnamefont {Loss}},\ }\bibfield
  {journal} {\bibinfo  {journal} {Physical Review B}\ }\textbf {\bibinfo
  {volume} {96}},\ \href {https://doi.org/10.1103/physrevb.96.224414}
  {10.1103/physrevb.96.224414} (\bibinfo {year} {2017})\BibitemShut {NoStop}%
\bibitem [{\citenamefont {Bonbien}\ \emph {et~al.}(2021)\citenamefont
  {Bonbien}, \citenamefont {Zhuo}, \citenamefont {Salimath}, \citenamefont
  {Ly}, \citenamefont {Abbout},\ and\ \citenamefont {Manchon}}]{Bonbien2021}%
  \BibitemOpen
  \bibfield  {author} {\bibinfo {author} {\bibfnamefont {V.}~\bibnamefont
  {Bonbien}}, \bibinfo {author} {\bibfnamefont {F.}~\bibnamefont {Zhuo}},
  \bibinfo {author} {\bibfnamefont {A.}~\bibnamefont {Salimath}}, \bibinfo
  {author} {\bibfnamefont {O.}~\bibnamefont {Ly}}, \bibinfo {author}
  {\bibfnamefont {A.}~\bibnamefont {Abbout}},\ and\ \bibinfo {author}
  {\bibfnamefont {A.}~\bibnamefont {Manchon}},\ }\href
  {https://doi.org/10.1088/1361-6463/ac28fa} {\bibfield  {journal} {\bibinfo
  {journal} {Journal of Physics D: Applied Physics}\ }\textbf {\bibinfo
  {volume} {55}},\ \bibinfo {pages} {103002} (\bibinfo {year}
  {2021})}\BibitemShut {NoStop}%
\bibitem [{\citenamefont {Šmejkal}\ \emph {et~al.}(2022)\citenamefont
  {Šmejkal}, \citenamefont {MacDonald}, \citenamefont {Sinova}, \citenamefont
  {Nakatsuji},\ and\ \citenamefont {Jungwirth}}]{Smejkal2022}%
  \BibitemOpen
  \bibfield  {author} {\bibinfo {author} {\bibfnamefont {L.}~\bibnamefont
  {Šmejkal}}, \bibinfo {author} {\bibfnamefont {A.~H.}\ \bibnamefont
  {MacDonald}}, \bibinfo {author} {\bibfnamefont {J.}~\bibnamefont {Sinova}},
  \bibinfo {author} {\bibfnamefont {S.}~\bibnamefont {Nakatsuji}},\ and\
  \bibinfo {author} {\bibfnamefont {T.}~\bibnamefont {Jungwirth}},\ }\href
  {https://doi.org/10.1038/s41578-022-00430-3} {\bibfield  {journal} {\bibinfo
  {journal} {Nature Reviews Materials}\ }\textbf {\bibinfo {volume} {7}},\
  \bibinfo {pages} {482} (\bibinfo {year} {2022})}\BibitemShut {NoStop}%
\bibitem [{\citenamefont {Georges}\ \emph {et~al.}(1996)\citenamefont
  {Georges}, \citenamefont {Kotliar}, \citenamefont {Krauth},\ and\
  \citenamefont {Rozenberg}}]{kotliar1}%
  \BibitemOpen
  \bibfield  {author} {\bibinfo {author} {\bibfnamefont {A.}~\bibnamefont
  {Georges}}, \bibinfo {author} {\bibfnamefont {G.}~\bibnamefont {Kotliar}},
  \bibinfo {author} {\bibfnamefont {W.}~\bibnamefont {Krauth}},\ and\ \bibinfo
  {author} {\bibfnamefont {M.~J.}\ \bibnamefont {Rozenberg}},\ }\href
  {https://doi.org/10.1103/RevModPhys.68.13} {\bibfield  {journal} {\bibinfo
  {journal} {Rev. Mod. Phys.}\ }\textbf {\bibinfo {volume} {68}},\ \bibinfo
  {pages} {13} (\bibinfo {year} {1996})}\BibitemShut {NoStop}%
\bibitem [{\citenamefont {Katsnelson}\ \emph {et~al.}(2008)\citenamefont
  {Katsnelson}, \citenamefont {Irkhin}, \citenamefont {Chioncel}, \citenamefont
  {Lichtenstein},\ and\ \citenamefont {de~Groot}}]{halfmet}%
  \BibitemOpen
  \bibfield  {author} {\bibinfo {author} {\bibfnamefont {M.~I.}\ \bibnamefont
  {Katsnelson}}, \bibinfo {author} {\bibfnamefont {V.~Y.}\ \bibnamefont
  {Irkhin}}, \bibinfo {author} {\bibfnamefont {L.}~\bibnamefont {Chioncel}},
  \bibinfo {author} {\bibfnamefont {A.~I.}\ \bibnamefont {Lichtenstein}},\ and\
  \bibinfo {author} {\bibfnamefont {R.~A.}\ \bibnamefont {de~Groot}},\ }\href
  {https://doi.org/10.1103/RevModPhys.80.315} {\bibfield  {journal} {\bibinfo
  {journal} {Rev. Mod. Phys.}\ }\textbf {\bibinfo {volume} {80}},\ \bibinfo
  {pages} {315} (\bibinfo {year} {2008})}\BibitemShut {NoStop}%
\bibitem [{\citenamefont {Rohringer}\ \emph {et~al.}(2018)\citenamefont
  {Rohringer}, \citenamefont {Hafermann}, \citenamefont {Toschi}, \citenamefont
  {Katanin}, \citenamefont {Antipov}, \citenamefont {Katsnelson}, \citenamefont
  {Lichtenstein}, \citenamefont {Rubtsov},\ and\ \citenamefont
  {Held}}]{beyond}%
  \BibitemOpen
  \bibfield  {author} {\bibinfo {author} {\bibfnamefont {G.}~\bibnamefont
  {Rohringer}}, \bibinfo {author} {\bibfnamefont {H.}~\bibnamefont
  {Hafermann}}, \bibinfo {author} {\bibfnamefont {A.}~\bibnamefont {Toschi}},
  \bibinfo {author} {\bibfnamefont {A.~A.}\ \bibnamefont {Katanin}}, \bibinfo
  {author} {\bibfnamefont {A.~E.}\ \bibnamefont {Antipov}}, \bibinfo {author}
  {\bibfnamefont {M.~I.}\ \bibnamefont {Katsnelson}}, \bibinfo {author}
  {\bibfnamefont {A.~I.}\ \bibnamefont {Lichtenstein}}, \bibinfo {author}
  {\bibfnamefont {A.~N.}\ \bibnamefont {Rubtsov}},\ and\ \bibinfo {author}
  {\bibfnamefont {K.}~\bibnamefont {Held}},\ }\href
  {https://doi.org/10.1103/RevModPhys.90.025003} {\bibfield  {journal}
  {\bibinfo  {journal} {Rev. Mod. Phys.}\ }\textbf {\bibinfo {volume} {90}},\
  \bibinfo {pages} {025003} (\bibinfo {year} {2018})}\BibitemShut {NoStop}%
\bibitem [{\citenamefont {Kim}\ \emph {et~al.}(2018)\citenamefont {Kim},
  \citenamefont {Kim}, \citenamefont {Sandilands}, \citenamefont {Sinn},
  \citenamefont {Lee}, \citenamefont {Son}, \citenamefont {Lee}, \citenamefont
  {Choi}, \citenamefont {Kim}, \citenamefont {Park}, \citenamefont {Jeon},
  \citenamefont {Kim}, \citenamefont {Park}, \citenamefont {Park},
  \citenamefont {Moon},\ and\ \citenamefont {Noh}}]{Kim2018}%
  \BibitemOpen
  \bibfield  {author} {\bibinfo {author} {\bibfnamefont {S.~Y.}\ \bibnamefont
  {Kim}}, \bibinfo {author} {\bibfnamefont {T.~Y.}\ \bibnamefont {Kim}},
  \bibinfo {author} {\bibfnamefont {L.~J.}\ \bibnamefont {Sandilands}},
  \bibinfo {author} {\bibfnamefont {S.}~\bibnamefont {Sinn}}, \bibinfo {author}
  {\bibfnamefont {M.-C.}\ \bibnamefont {Lee}}, \bibinfo {author} {\bibfnamefont
  {J.}~\bibnamefont {Son}}, \bibinfo {author} {\bibfnamefont {S.}~\bibnamefont
  {Lee}}, \bibinfo {author} {\bibfnamefont {K.-Y.}\ \bibnamefont {Choi}},
  \bibinfo {author} {\bibfnamefont {W.}~\bibnamefont {Kim}}, \bibinfo {author}
  {\bibfnamefont {B.-G.}\ \bibnamefont {Park}}, \bibinfo {author}
  {\bibfnamefont {C.}~\bibnamefont {Jeon}}, \bibinfo {author} {\bibfnamefont
  {H.-D.}\ \bibnamefont {Kim}}, \bibinfo {author} {\bibfnamefont {C.-H.}\
  \bibnamefont {Park}}, \bibinfo {author} {\bibfnamefont {J.-G.}\ \bibnamefont
  {Park}}, \bibinfo {author} {\bibfnamefont {S.}~\bibnamefont {Moon}},\ and\
  \bibinfo {author} {\bibfnamefont {T.}~\bibnamefont {Noh}},\ }\bibfield
  {journal} {\bibinfo  {journal} {Physical Review Letters}\ }\textbf {\bibinfo
  {volume} {120}},\ \href {https://doi.org/10.1103/physrevlett.120.136402}
  {10.1103/physrevlett.120.136402} (\bibinfo {year} {2018})\BibitemShut
  {NoStop}%
\bibitem [{\citenamefont {Acharya}\ \emph {et~al.}(2023)\citenamefont
  {Acharya}, \citenamefont {Pashov}, \citenamefont {Weber}, \citenamefont {van
  Schilfgaarde}, \citenamefont {Lichtenstein},\ and\ \citenamefont
  {Katsnelson}}]{Acharya2023}%
  \BibitemOpen
  \bibfield  {author} {\bibinfo {author} {\bibfnamefont {S.}~\bibnamefont
  {Acharya}}, \bibinfo {author} {\bibfnamefont {D.}~\bibnamefont {Pashov}},
  \bibinfo {author} {\bibfnamefont {C.}~\bibnamefont {Weber}}, \bibinfo
  {author} {\bibfnamefont {M.}~\bibnamefont {van Schilfgaarde}}, \bibinfo
  {author} {\bibfnamefont {A.~I.}\ \bibnamefont {Lichtenstein}},\ and\ \bibinfo
  {author} {\bibfnamefont {M.~I.}\ \bibnamefont {Katsnelson}},\ }\bibfield
  {journal} {\bibinfo  {journal} {Nature Communications}\ }\textbf {\bibinfo
  {volume} {14}},\ \href {https://doi.org/10.1038/s41467-023-41314-6}
  {10.1038/s41467-023-41314-6} (\bibinfo {year} {2023})\BibitemShut {NoStop}%
\bibitem [{\citenamefont {Edwards}\ and\ \citenamefont
  {Hertz}(1973)}]{Edwards1973}%
  \BibitemOpen
  \bibfield  {author} {\bibinfo {author} {\bibfnamefont {D.~M.}\ \bibnamefont
  {Edwards}}\ and\ \bibinfo {author} {\bibfnamefont {J.~A.}\ \bibnamefont
  {Hertz}},\ }\href {https://doi.org/10.1088/0305-4608/3/12/019} {\bibfield
  {journal} {\bibinfo  {journal} {J. Phys. F}\ }\textbf {\bibinfo {volume}
  {3}},\ \bibinfo {pages} {2191} (\bibinfo {year} {1973})}\BibitemShut
  {NoStop}%
\bibitem [{\citenamefont {Hertz}\ and\ \citenamefont
  {Edwards}(1973)}]{Hertz1973}%
  \BibitemOpen
  \bibfield  {author} {\bibinfo {author} {\bibfnamefont {J.~A.}\ \bibnamefont
  {Hertz}}\ and\ \bibinfo {author} {\bibfnamefont {D.~M.}\ \bibnamefont
  {Edwards}},\ }\href {https://doi.org/10.1088/0305-4608/3/12/018} {\bibfield
  {journal} {\bibinfo  {journal} {J. Phys. F}\ }\textbf {\bibinfo {volume}
  {3}},\ \bibinfo {pages} {2174} (\bibinfo {year} {1973})}\BibitemShut
  {NoStop}%
\bibitem [{\citenamefont {Irkhin}\ and\ \citenamefont
  {Katsnelson}(2000)}]{IK_AFM}%
  \BibitemOpen
  \bibfield  {author} {\bibinfo {author} {\bibfnamefont {V.~Y.}\ \bibnamefont
  {Irkhin}}\ and\ \bibinfo {author} {\bibfnamefont {M.~I.}\ \bibnamefont
  {Katsnelson}},\ }\href {https://doi.org/10.1103/PhysRevB.62.5647} {\bibfield
  {journal} {\bibinfo  {journal} {Phys. Rev. B}\ }\textbf {\bibinfo {volume}
  {62}},\ \bibinfo {pages} {5647} (\bibinfo {year} {2000})}\BibitemShut
  {NoStop}%
\bibitem [{\citenamefont {Demler}\ \emph {et~al.}(2004)\citenamefont {Demler},
  \citenamefont {Hanke},\ and\ \citenamefont {Zhang}}]{cuprates1}%
  \BibitemOpen
  \bibfield  {author} {\bibinfo {author} {\bibfnamefont {E.}~\bibnamefont
  {Demler}}, \bibinfo {author} {\bibfnamefont {W.}~\bibnamefont {Hanke}},\ and\
  \bibinfo {author} {\bibfnamefont {S.-C.}\ \bibnamefont {Zhang}},\ }\href
  {https://doi.org/10.1103/RevModPhys.76.909} {\bibfield  {journal} {\bibinfo
  {journal} {Rev. Mod. Phys.}\ }\textbf {\bibinfo {volume} {76}},\ \bibinfo
  {pages} {909} (\bibinfo {year} {2004})}\BibitemShut {NoStop}%
\bibitem [{\citenamefont {Lee}\ \emph {et~al.}(2006)\citenamefont {Lee},
  \citenamefont {Nagaosa},\ and\ \citenamefont {Wen}}]{cuprates2}%
  \BibitemOpen
  \bibfield  {author} {\bibinfo {author} {\bibfnamefont {P.~A.}\ \bibnamefont
  {Lee}}, \bibinfo {author} {\bibfnamefont {N.}~\bibnamefont {Nagaosa}},\ and\
  \bibinfo {author} {\bibfnamefont {X.-G.}\ \bibnamefont {Wen}},\ }\href
  {https://doi.org/10.1103/RevModPhys.78.17} {\bibfield  {journal} {\bibinfo
  {journal} {Rev. Mod. Phys.}\ }\textbf {\bibinfo {volume} {78}},\ \bibinfo
  {pages} {17} (\bibinfo {year} {2006})}\BibitemShut {NoStop}%
\bibitem [{\citenamefont {Scalapino}(2012)}]{cuprates3}%
  \BibitemOpen
  \bibfield  {author} {\bibinfo {author} {\bibfnamefont {D.~J.}\ \bibnamefont
  {Scalapino}},\ }\href {https://doi.org/10.1103/RevModPhys.84.1383} {\bibfield
   {journal} {\bibinfo  {journal} {Rev. Mod. Phys.}\ }\textbf {\bibinfo
  {volume} {84}},\ \bibinfo {pages} {1383} (\bibinfo {year}
  {2012})}\BibitemShut {NoStop}%
\bibitem [{\citenamefont {Dai}(2015)}]{pnictides}%
  \BibitemOpen
  \bibfield  {author} {\bibinfo {author} {\bibfnamefont {P.}~\bibnamefont
  {Dai}},\ }\href {https://doi.org/10.1103/RevModPhys.87.855} {\bibfield
  {journal} {\bibinfo  {journal} {Rev. Mod. Phys.}\ }\textbf {\bibinfo {volume}
  {87}},\ \bibinfo {pages} {855} (\bibinfo {year} {2015})}\BibitemShut
  {NoStop}%
\bibitem [{\citenamefont {Anisimov}\ \emph {et~al.}(1997)\citenamefont
  {Anisimov}, \citenamefont {Poteryaev}, \citenamefont {Korotin}, \citenamefont
  {Anokhin},\ and\ \citenamefont {Kotliar}}]{Anisimov1997}%
  \BibitemOpen
  \bibfield  {author} {\bibinfo {author} {\bibfnamefont {V.~I.}\ \bibnamefont
  {Anisimov}}, \bibinfo {author} {\bibfnamefont {A.~I.}\ \bibnamefont
  {Poteryaev}}, \bibinfo {author} {\bibfnamefont {M.~A.}\ \bibnamefont
  {Korotin}}, \bibinfo {author} {\bibfnamefont {A.~O.}\ \bibnamefont
  {Anokhin}},\ and\ \bibinfo {author} {\bibfnamefont {G.}~\bibnamefont
  {Kotliar}},\ }\href {https://doi.org/10.1088/0953-8984/9/35/010} {\bibfield
  {journal} {\bibinfo  {journal} {J. Phys.: Condens. Mat.}\ }\textbf {\bibinfo
  {volume} {9}},\ \bibinfo {pages} {7359} (\bibinfo {year} {1997})}\BibitemShut
  {NoStop}%
\bibitem [{\citenamefont {Lichtenstein}\ and\ \citenamefont
  {Katsnelson}(1998)}]{ldaplusplus}%
  \BibitemOpen
  \bibfield  {author} {\bibinfo {author} {\bibfnamefont {A.~I.}\ \bibnamefont
  {Lichtenstein}}\ and\ \bibinfo {author} {\bibfnamefont {M.~I.}\ \bibnamefont
  {Katsnelson}},\ }\href {https://doi.org/10.1103/PhysRevB.57.6884} {\bibfield
  {journal} {\bibinfo  {journal} {Phys. Rev. B}\ }\textbf {\bibinfo {volume}
  {57}},\ \bibinfo {pages} {6884} (\bibinfo {year} {1998})}\BibitemShut
  {NoStop}%
\bibitem [{\citenamefont {Kotliar}\ \emph {et~al.}(2006)\citenamefont
  {Kotliar}, \citenamefont {Savrasov}, \citenamefont {Haule}, \citenamefont
  {Oudovenko}, \citenamefont {Parcollet},\ and\ \citenamefont
  {Marianetti}}]{Kotliar2}%
  \BibitemOpen
  \bibfield  {author} {\bibinfo {author} {\bibfnamefont {G.}~\bibnamefont
  {Kotliar}}, \bibinfo {author} {\bibfnamefont {S.~Y.}\ \bibnamefont
  {Savrasov}}, \bibinfo {author} {\bibfnamefont {K.}~\bibnamefont {Haule}},
  \bibinfo {author} {\bibfnamefont {V.~S.}\ \bibnamefont {Oudovenko}}, \bibinfo
  {author} {\bibfnamefont {O.}~\bibnamefont {Parcollet}},\ and\ \bibinfo
  {author} {\bibfnamefont {C.~A.}\ \bibnamefont {Marianetti}},\ }\href
  {https://doi.org/10.1103/RevModPhys.78.865} {\bibfield  {journal} {\bibinfo
  {journal} {Rev. Mod. Phys.}\ }\textbf {\bibinfo {volume} {78}},\ \bibinfo
  {pages} {865} (\bibinfo {year} {2006})}\BibitemShut {NoStop}%
\bibitem [{\citenamefont {Lichtenstein}\ and\ \citenamefont
  {Katsnelson}(2000)}]{Lichtenstein2000}%
  \BibitemOpen
  \bibfield  {author} {\bibinfo {author} {\bibfnamefont {A.~I.}\ \bibnamefont
  {Lichtenstein}}\ and\ \bibinfo {author} {\bibfnamefont {M.~I.}\ \bibnamefont
  {Katsnelson}},\ }\href {https://doi.org/10.1103/physrevb.62.r9283} {\bibfield
   {journal} {\bibinfo  {journal} {Physical Review B}\ }\textbf {\bibinfo
  {volume} {62}},\ \bibinfo {pages} {R9283–R9286} (\bibinfo {year}
  {2000})}\BibitemShut {NoStop}%
\bibitem [{\citenamefont {Sangiovanni}\ \emph {et~al.}(2006)\citenamefont
  {Sangiovanni}, \citenamefont {Gunnarsson}, \citenamefont {Koch},
  \citenamefont {Castellani},\ and\ \citenamefont {Capone}}]{Sangiovanni2006}%
  \BibitemOpen
  \bibfield  {author} {\bibinfo {author} {\bibfnamefont {G.}~\bibnamefont
  {Sangiovanni}}, \bibinfo {author} {\bibfnamefont {O.}~\bibnamefont
  {Gunnarsson}}, \bibinfo {author} {\bibfnamefont {E.}~\bibnamefont {Koch}},
  \bibinfo {author} {\bibfnamefont {C.}~\bibnamefont {Castellani}},\ and\
  \bibinfo {author} {\bibfnamefont {M.}~\bibnamefont {Capone}},\ }\bibfield
  {journal} {\bibinfo  {journal} {Physical Review Letters}\ }\textbf {\bibinfo
  {volume} {97}},\ \href {https://doi.org/10.1103/physrevlett.97.046404}
  {10.1103/physrevlett.97.046404} (\bibinfo {year} {2006})\BibitemShut
  {NoStop}%
\bibitem [{\citenamefont {Miura}\ and\ \citenamefont
  {Fujiwara}(2008)}]{Miura2008}%
  \BibitemOpen
  \bibfield  {author} {\bibinfo {author} {\bibfnamefont {O.}~\bibnamefont
  {Miura}}\ and\ \bibinfo {author} {\bibfnamefont {T.}~\bibnamefont
  {Fujiwara}},\ }\bibfield  {journal} {\bibinfo  {journal} {Physical Review B}\
  }\textbf {\bibinfo {volume} {77}},\ \href
  {https://doi.org/10.1103/physrevb.77.195124} {10.1103/physrevb.77.195124}
  (\bibinfo {year} {2008})\BibitemShut {NoStop}%
\bibitem [{\citenamefont {Rubtsov}\ \emph {et~al.}(2008)\citenamefont
  {Rubtsov}, \citenamefont {Katsnelson},\ and\ \citenamefont
  {Lichtenstein}}]{Rubtsov2008}%
  \BibitemOpen
  \bibfield  {author} {\bibinfo {author} {\bibfnamefont {A.~N.}\ \bibnamefont
  {Rubtsov}}, \bibinfo {author} {\bibfnamefont {M.~I.}\ \bibnamefont
  {Katsnelson}},\ and\ \bibinfo {author} {\bibfnamefont {A.~I.}\ \bibnamefont
  {Lichtenstein}},\ }\bibfield  {journal} {\bibinfo  {journal} {Physical Review
  B}\ }\textbf {\bibinfo {volume} {77}},\ \href
  {https://doi.org/10.1103/physrevb.77.033101} {10.1103/physrevb.77.033101}
  (\bibinfo {year} {2008})\BibitemShut {NoStop}%
\bibitem [{\citenamefont {Rubtsov}\ \emph {et~al.}(2009)\citenamefont
  {Rubtsov}, \citenamefont {Katsnelson}, \citenamefont {Lichtenstein},\ and\
  \citenamefont {Georges}}]{Rubtsov2009}%
  \BibitemOpen
  \bibfield  {author} {\bibinfo {author} {\bibfnamefont {A.~N.}\ \bibnamefont
  {Rubtsov}}, \bibinfo {author} {\bibfnamefont {M.~I.}\ \bibnamefont
  {Katsnelson}}, \bibinfo {author} {\bibfnamefont {A.~I.}\ \bibnamefont
  {Lichtenstein}},\ and\ \bibinfo {author} {\bibfnamefont {A.}~\bibnamefont
  {Georges}},\ }\bibfield  {journal} {\bibinfo  {journal} {Physical Review B}\
  }\textbf {\bibinfo {volume} {79}},\ \href
  {https://doi.org/10.1103/physrevb.79.045133} {10.1103/physrevb.79.045133}
  (\bibinfo {year} {2009})\BibitemShut {NoStop}%
\bibitem [{\citenamefont {Hedin}(1965)}]{Hedin1965}%
  \BibitemOpen
  \bibfield  {author} {\bibinfo {author} {\bibfnamefont {L.}~\bibnamefont
  {Hedin}},\ }\href {https://doi.org/10.1103/physrev.139.a796} {\bibfield
  {journal} {\bibinfo  {journal} {Phys. Rev.}\ }\textbf {\bibinfo {volume}
  {139}},\ \bibinfo {pages} {A796} (\bibinfo {year} {1965})}\BibitemShut
  {NoStop}%
\bibitem [{\citenamefont {Müller}\ \emph {et~al.}(2016)\citenamefont
  {Müller}, \citenamefont {Friedrich},\ and\ \citenamefont
  {Blügel}}]{Mueller2016}%
  \BibitemOpen
  \bibfield  {author} {\bibinfo {author} {\bibfnamefont {M.~C. T.~D.}\
  \bibnamefont {Müller}}, \bibinfo {author} {\bibfnamefont {C.}~\bibnamefont
  {Friedrich}},\ and\ \bibinfo {author} {\bibfnamefont {S.}~\bibnamefont
  {Blügel}},\ }\bibfield  {journal} {\bibinfo  {journal} {Phys. Rev. B}\
  }\textbf {\bibinfo {volume} {94}},\ \href
  {https://doi.org/10.1103/physrevb.94.064433} {10.1103/physrevb.94.064433}
  (\bibinfo {year} {2016})\BibitemShut {NoStop}%
\bibitem [{\citenamefont {Nabok}\ \emph {et~al.}(2021)\citenamefont {Nabok},
  \citenamefont {Blügel},\ and\ \citenamefont {Friedrich}}]{Nabok2021}%
  \BibitemOpen
  \bibfield  {author} {\bibinfo {author} {\bibfnamefont {D.}~\bibnamefont
  {Nabok}}, \bibinfo {author} {\bibfnamefont {S.}~\bibnamefont {Blügel}},\
  and\ \bibinfo {author} {\bibfnamefont {C.}~\bibnamefont {Friedrich}},\
  }\bibfield  {journal} {\bibinfo  {journal} {npj Comp. Mat.}\ }\textbf
  {\bibinfo {volume} {7}},\ \href {https://doi.org/10.1038/s41524-021-00649-8}
  {10.1038/s41524-021-00649-8} (\bibinfo {year} {2021})\BibitemShut {NoStop}%
\bibitem [{\citenamefont {Buczek}\ \emph {et~al.}(2011)\citenamefont {Buczek},
  \citenamefont {Ernst},\ and\ \citenamefont {Sandratskii}}]{Buczek2011}%
  \BibitemOpen
  \bibfield  {author} {\bibinfo {author} {\bibfnamefont {P.}~\bibnamefont
  {Buczek}}, \bibinfo {author} {\bibfnamefont {A.}~\bibnamefont {Ernst}},\ and\
  \bibinfo {author} {\bibfnamefont {L.~M.}\ \bibnamefont {Sandratskii}},\
  }\bibfield  {journal} {\bibinfo  {journal} {Phys. Rev. B}\ }\textbf {\bibinfo
  {volume} {84}},\ \href {https://doi.org/10.1103/physrevb.84.174418}
  {10.1103/physrevb.84.174418} (\bibinfo {year} {2011})\BibitemShut {NoStop}%
\bibitem [{\citenamefont {Hoffmann}\ \emph {et~al.}(2020)\citenamefont
  {Hoffmann}, \citenamefont {Ernst}, \citenamefont {Hergert}, \citenamefont
  {Antonov}, \citenamefont {Adeagbo}, \citenamefont {Geilhufe},\ and\
  \citenamefont {Hamed}}]{Hoffmann2020}%
  \BibitemOpen
  \bibfield  {author} {\bibinfo {author} {\bibfnamefont {M.}~\bibnamefont
  {Hoffmann}}, \bibinfo {author} {\bibfnamefont {A.}~\bibnamefont {Ernst}},
  \bibinfo {author} {\bibfnamefont {W.}~\bibnamefont {Hergert}}, \bibinfo
  {author} {\bibfnamefont {V.~N.}\ \bibnamefont {Antonov}}, \bibinfo {author}
  {\bibfnamefont {W.~A.}\ \bibnamefont {Adeagbo}}, \bibinfo {author}
  {\bibfnamefont {R.~M.}\ \bibnamefont {Geilhufe}},\ and\ \bibinfo {author}
  {\bibfnamefont {H.~B.}\ \bibnamefont {Hamed}},\ }\bibfield  {journal}
  {\bibinfo  {journal} {physica status solidi (b)}\ }\textbf {\bibinfo {volume}
  {257}},\ \href {https://doi.org/10.1002/pssb.201900671}
  {10.1002/pssb.201900671} (\bibinfo {year} {2020})\BibitemShut {NoStop}%
\bibitem [{\citenamefont {Gross}\ and\ \citenamefont {Kohn}(1990)}]{Gross1990}%
  \BibitemOpen
  \bibfield  {author} {\bibinfo {author} {\bibfnamefont {E.}~\bibnamefont
  {Gross}}\ and\ \bibinfo {author} {\bibfnamefont {W.}~\bibnamefont {Kohn}},\
  }\bibinfo {title} {Time-dependent density-functional theory},\ in\ \href
  {https://doi.org/10.1016/s0065-3276(08)60600-0} {\emph {\bibinfo {booktitle}
  {Density Functional Theory of Many-Fermion Systems}}}\ (\bibinfo  {publisher}
  {Elsevier},\ \bibinfo {year} {1990})\ pp.\ \bibinfo {pages}
  {255--291}\BibitemShut {NoStop}%
\bibitem [{\citenamefont {Sandratskii}\ and\ \citenamefont
  {Buczek}(2012)}]{Sandratskii2012}%
  \BibitemOpen
  \bibfield  {author} {\bibinfo {author} {\bibfnamefont {L.~M.}\ \bibnamefont
  {Sandratskii}}\ and\ \bibinfo {author} {\bibfnamefont {P.}~\bibnamefont
  {Buczek}},\ }\bibfield  {journal} {\bibinfo  {journal} {Phys. Rev. B}\
  }\textbf {\bibinfo {volume} {85}},\ \href
  {https://doi.org/10.1103/physrevb.85.020406} {10.1103/physrevb.85.020406}
  (\bibinfo {year} {2012})\BibitemShut {NoStop}%
\bibitem [{\citenamefont {Vignale}\ and\ \citenamefont
  {Singwi}(1985)}]{Vignale1985b}%
  \BibitemOpen
  \bibfield  {author} {\bibinfo {author} {\bibfnamefont {G.}~\bibnamefont
  {Vignale}}\ and\ \bibinfo {author} {\bibfnamefont {K.~S.}\ \bibnamefont
  {Singwi}},\ }\href {https://doi.org/10.1103/physrevb.32.2156} {\bibfield
  {journal} {\bibinfo  {journal} {Phys. Rev. B}\ }\textbf {\bibinfo {volume}
  {32}},\ \bibinfo {pages} {2156} (\bibinfo {year} {1985})}\BibitemShut
  {NoStop}%
\bibitem [{\citenamefont {Ng}\ and\ \citenamefont {Singwi}(1986)}]{Ng1986}%
  \BibitemOpen
  \bibfield  {author} {\bibinfo {author} {\bibfnamefont {T.~K.}\ \bibnamefont
  {Ng}}\ and\ \bibinfo {author} {\bibfnamefont {K.~S.}\ \bibnamefont
  {Singwi}},\ }\href {https://doi.org/10.1103/physrevb.34.7738} {\bibfield
  {journal} {\bibinfo  {journal} {Phys. Rev. B}\ }\textbf {\bibinfo {volume}
  {34}},\ \bibinfo {pages} {7738} (\bibinfo {year} {1986})}\BibitemShut
  {NoStop}%
\bibitem [{\citenamefont {Müller}\ \emph {et~al.}(2019)\citenamefont
  {Müller}, \citenamefont {Blügel},\ and\ \citenamefont
  {Friedrich}}]{Mueller2019}%
  \BibitemOpen
  \bibfield  {author} {\bibinfo {author} {\bibfnamefont {M.~C. T.~D.}\
  \bibnamefont {Müller}}, \bibinfo {author} {\bibfnamefont {S.}~\bibnamefont
  {Blügel}},\ and\ \bibinfo {author} {\bibfnamefont {C.}~\bibnamefont
  {Friedrich}},\ }\bibfield  {journal} {\bibinfo  {journal} {Phys. Rev. B}\
  }\textbf {\bibinfo {volume} {100}},\ \href
  {https://doi.org/10.1103/physrevb.100.045130} {10.1103/physrevb.100.045130}
  (\bibinfo {year} {2019})\BibitemShut {NoStop}%
\bibitem [{\citenamefont {Buczek}\ \emph {et~al.}(2020)\citenamefont {Buczek},
  \citenamefont {Buczek}, \citenamefont {Vignale},\ and\ \citenamefont
  {Ernst}}]{Buczek2020}%
  \BibitemOpen
  \bibfield  {author} {\bibinfo {author} {\bibfnamefont {P.}~\bibnamefont
  {Buczek}}, \bibinfo {author} {\bibfnamefont {N.}~\bibnamefont {Buczek}},
  \bibinfo {author} {\bibfnamefont {G.}~\bibnamefont {Vignale}},\ and\ \bibinfo
  {author} {\bibfnamefont {A.}~\bibnamefont {Ernst}},\ }\bibfield  {journal}
  {\bibinfo  {journal} {Phys. Rev. B}\ }\textbf {\bibinfo {volume} {101}},\
  \href {https://doi.org/10.1103/physrevb.101.214420}
  {10.1103/physrevb.101.214420} (\bibinfo {year} {2020})\BibitemShut {NoStop}%
\bibitem [{\citenamefont {Lichtenstein}\ \emph {et~al.}(1987)\citenamefont
  {Lichtenstein}, \citenamefont {Katsnelson}, \citenamefont {Antropov},\ and\
  \citenamefont {Gubanov}}]{Lichtenstein1987}%
  \BibitemOpen
  \bibfield  {author} {\bibinfo {author} {\bibfnamefont {A.}~\bibnamefont
  {Lichtenstein}}, \bibinfo {author} {\bibfnamefont {M.}~\bibnamefont
  {Katsnelson}}, \bibinfo {author} {\bibfnamefont {V.}~\bibnamefont
  {Antropov}},\ and\ \bibinfo {author} {\bibfnamefont {V.}~\bibnamefont
  {Gubanov}},\ }\href {https://doi.org/10.1016/0304-8853(87)90721-9} {\bibfield
   {journal} {\bibinfo  {journal} {J. Magn. Magn. Mater.}\ }\textbf {\bibinfo
  {volume} {67}},\ \bibinfo {pages} {65} (\bibinfo {year} {1987})}\BibitemShut
  {NoStop}%
\bibitem [{\citenamefont {S{\'{a}}nchez-Barriga}\ \emph
  {et~al.}(2012)\citenamefont {S{\'{a}}nchez-Barriga}, \citenamefont {Braun},
  \citenamefont {Min{\'{a}}r}, \citenamefont {Marco}, \citenamefont
  {Varykhalov}, \citenamefont {Rader}, \citenamefont {Boni}, \citenamefont
  {Bellini}, \citenamefont {Manghi}, \citenamefont {Ebert}, \citenamefont
  {Katsnelson}, \citenamefont {Lichtenstein}, \citenamefont {Eriksson},
  \citenamefont {Eberhardt}, \citenamefont {Dürr},\ and\ \citenamefont
  {Fink}}]{SanchezBarriga2012}%
  \BibitemOpen
  \bibfield  {author} {\bibinfo {author} {\bibfnamefont {J.}~\bibnamefont
  {S{\'{a}}nchez-Barriga}}, \bibinfo {author} {\bibfnamefont {J.}~\bibnamefont
  {Braun}}, \bibinfo {author} {\bibfnamefont {J.}~\bibnamefont {Min{\'{a}}r}},
  \bibinfo {author} {\bibfnamefont {I.~D.}\ \bibnamefont {Marco}}, \bibinfo
  {author} {\bibfnamefont {A.}~\bibnamefont {Varykhalov}}, \bibinfo {author}
  {\bibfnamefont {O.}~\bibnamefont {Rader}}, \bibinfo {author} {\bibfnamefont
  {V.}~\bibnamefont {Boni}}, \bibinfo {author} {\bibfnamefont {V.}~\bibnamefont
  {Bellini}}, \bibinfo {author} {\bibfnamefont {F.}~\bibnamefont {Manghi}},
  \bibinfo {author} {\bibfnamefont {H.}~\bibnamefont {Ebert}}, \bibinfo
  {author} {\bibfnamefont {M.~I.}\ \bibnamefont {Katsnelson}}, \bibinfo
  {author} {\bibfnamefont {A.~I.}\ \bibnamefont {Lichtenstein}}, \bibinfo
  {author} {\bibfnamefont {O.}~\bibnamefont {Eriksson}}, \bibinfo {author}
  {\bibfnamefont {W.}~\bibnamefont {Eberhardt}}, \bibinfo {author}
  {\bibfnamefont {H.~A.}\ \bibnamefont {Dürr}},\ and\ \bibinfo {author}
  {\bibfnamefont {J.}~\bibnamefont {Fink}},\ }\bibfield  {journal} {\bibinfo
  {journal} {Phys. Rev. B}\ }\textbf {\bibinfo {volume} {85}},\ \href
  {https://doi.org/10.1103/physrevb.85.205109} {10.1103/physrevb.85.205109}
  (\bibinfo {year} {2012})\BibitemShut {NoStop}%
\bibitem [{\citenamefont {Park}\ \emph {et~al.}(2020)\citenamefont {Park},
  \citenamefont {Kwon},\ and\ \citenamefont {Lake}}]{Park2020}%
  \BibitemOpen
  \bibfield  {author} {\bibinfo {author} {\bibfnamefont {I.~J.}\ \bibnamefont
  {Park}}, \bibinfo {author} {\bibfnamefont {S.}~\bibnamefont {Kwon}},\ and\
  \bibinfo {author} {\bibfnamefont {R.~K.}\ \bibnamefont {Lake}},\ }\bibfield
  {journal} {\bibinfo  {journal} {Physical Review B}\ }\textbf {\bibinfo
  {volume} {102}},\ \href {https://doi.org/10.1103/physrevb.102.224426}
  {10.1103/physrevb.102.224426} (\bibinfo {year} {2020})\BibitemShut {NoStop}%
\bibitem [{\citenamefont {Reimers}\ \emph {et~al.}(2024)\citenamefont
  {Reimers}, \citenamefont {Odenbreit}, \citenamefont {Šmejkal}, \citenamefont
  {Strocov}, \citenamefont {Constantinou}, \citenamefont {Hellenes},
  \citenamefont {Jaeschke~Ubiergo}, \citenamefont {Campos}, \citenamefont
  {Bharadwaj}, \citenamefont {Chakraborty}, \citenamefont {Denneulin},
  \citenamefont {Shi}, \citenamefont {Dunin-Borkowski}, \citenamefont {Das},
  \citenamefont {Kläui}, \citenamefont {Sinova},\ and\ \citenamefont
  {Jourdan}}]{Reimers2024}%
  \BibitemOpen
  \bibfield  {author} {\bibinfo {author} {\bibfnamefont {S.}~\bibnamefont
  {Reimers}}, \bibinfo {author} {\bibfnamefont {L.}~\bibnamefont {Odenbreit}},
  \bibinfo {author} {\bibfnamefont {L.}~\bibnamefont {Šmejkal}}, \bibinfo
  {author} {\bibfnamefont {V.~N.}\ \bibnamefont {Strocov}}, \bibinfo {author}
  {\bibfnamefont {P.}~\bibnamefont {Constantinou}}, \bibinfo {author}
  {\bibfnamefont {A.~B.}\ \bibnamefont {Hellenes}}, \bibinfo {author}
  {\bibfnamefont {R.}~\bibnamefont {Jaeschke~Ubiergo}}, \bibinfo {author}
  {\bibfnamefont {W.~H.}\ \bibnamefont {Campos}}, \bibinfo {author}
  {\bibfnamefont {V.~K.}\ \bibnamefont {Bharadwaj}}, \bibinfo {author}
  {\bibfnamefont {A.}~\bibnamefont {Chakraborty}}, \bibinfo {author}
  {\bibfnamefont {T.}~\bibnamefont {Denneulin}}, \bibinfo {author}
  {\bibfnamefont {W.}~\bibnamefont {Shi}}, \bibinfo {author} {\bibfnamefont
  {R.~E.}\ \bibnamefont {Dunin-Borkowski}}, \bibinfo {author} {\bibfnamefont
  {S.}~\bibnamefont {Das}}, \bibinfo {author} {\bibfnamefont {M.}~\bibnamefont
  {Kläui}}, \bibinfo {author} {\bibfnamefont {J.}~\bibnamefont {Sinova}},\
  and\ \bibinfo {author} {\bibfnamefont {M.}~\bibnamefont {Jourdan}},\
  }\bibfield  {journal} {\bibinfo  {journal} {Nature Communications}\ }\textbf
  {\bibinfo {volume} {15}},\ \href {https://doi.org/10.1038/s41467-024-46476-5}
  {10.1038/s41467-024-46476-5} (\bibinfo {year} {2024})\BibitemShut {NoStop}%
\bibitem [{\citenamefont {Dijkstra}\ \emph {et~al.}(1989)\citenamefont
  {Dijkstra}, \citenamefont {Bruggen}, \citenamefont {Haas},\ and\
  \citenamefont {Groot}}]{Dijkstra1989}%
  \BibitemOpen
  \bibfield  {author} {\bibinfo {author} {\bibfnamefont {J.}~\bibnamefont
  {Dijkstra}}, \bibinfo {author} {\bibfnamefont {C.~F.~v.}\ \bibnamefont
  {Bruggen}}, \bibinfo {author} {\bibfnamefont {C.}~\bibnamefont {Haas}},\ and\
  \bibinfo {author} {\bibfnamefont {R.~A.~d.}\ \bibnamefont {Groot}},\ }\href
  {https://doi.org/10.1088/0953-8984/1/46/009} {\bibfield  {journal} {\bibinfo
  {journal} {Journal of Physics: Condensed Matter}\ }\textbf {\bibinfo {volume}
  {1}},\ \bibinfo {pages} {9163} (\bibinfo {year} {1989})}\BibitemShut
  {NoStop}%
\bibitem [{\citenamefont {Ito}\ \emph {et~al.}(2007)\citenamefont {Ito},
  \citenamefont {Ido},\ and\ \citenamefont {Motizuki}}]{Ito2007}%
  \BibitemOpen
  \bibfield  {author} {\bibinfo {author} {\bibfnamefont {T.}~\bibnamefont
  {Ito}}, \bibinfo {author} {\bibfnamefont {H.}~\bibnamefont {Ido}},\ and\
  \bibinfo {author} {\bibfnamefont {K.}~\bibnamefont {Motizuki}},\ }\href
  {https://doi.org/10.1016/j.jmmm.2006.10.470} {\bibfield  {journal} {\bibinfo
  {journal} {Journal of Magnetism and Magnetic Materials}\ }\textbf {\bibinfo
  {volume} {310}},\ \bibinfo {pages} {e558–e559} (\bibinfo {year}
  {2007})}\BibitemShut {NoStop}%
\bibitem [{\citenamefont {Kahal}\ \emph {et~al.}(2007)\citenamefont {Kahal},
  \citenamefont {Zaoui},\ and\ \citenamefont {Ferhat}}]{Kahal2007}%
  \BibitemOpen
  \bibfield  {author} {\bibinfo {author} {\bibfnamefont {L.}~\bibnamefont
  {Kahal}}, \bibinfo {author} {\bibfnamefont {A.}~\bibnamefont {Zaoui}},\ and\
  \bibinfo {author} {\bibfnamefont {M.}~\bibnamefont {Ferhat}},\ }\bibfield
  {journal} {\bibinfo  {journal} {Journal of Applied Physics}\ }\textbf
  {\bibinfo {volume} {101}},\ \href {https://doi.org/10.1063/1.2732502}
  {10.1063/1.2732502} (\bibinfo {year} {2007})\BibitemShut {NoStop}%
\bibitem [{\citenamefont {Polesya}\ \emph {et~al.}(2011)\citenamefont
  {Polesya}, \citenamefont {Kuhn}, \citenamefont {Mankovsky}, \citenamefont
  {Ebert}, \citenamefont {Regus},\ and\ \citenamefont {Bensch}}]{Polesya2011}%
  \BibitemOpen
  \bibfield  {author} {\bibinfo {author} {\bibfnamefont {S.}~\bibnamefont
  {Polesya}}, \bibinfo {author} {\bibfnamefont {G.}~\bibnamefont {Kuhn}},
  \bibinfo {author} {\bibfnamefont {S.}~\bibnamefont {Mankovsky}}, \bibinfo
  {author} {\bibfnamefont {H.}~\bibnamefont {Ebert}}, \bibinfo {author}
  {\bibfnamefont {M.}~\bibnamefont {Regus}},\ and\ \bibinfo {author}
  {\bibfnamefont {W.}~\bibnamefont {Bensch}},\ }\href
  {https://doi.org/10.1088/0953-8984/24/3/036004} {\bibfield  {journal}
  {\bibinfo  {journal} {Journal of Physics: Condensed Matter}\ }\textbf
  {\bibinfo {volume} {24}},\ \bibinfo {pages} {036004} (\bibinfo {year}
  {2011})}\BibitemShut {NoStop}%
\bibitem [{\citenamefont {Snow}(1952)}]{Snow1952}%
  \BibitemOpen
  \bibfield  {author} {\bibinfo {author} {\bibfnamefont {A.~I.}\ \bibnamefont
  {Snow}},\ }\href {https://doi.org/10.1103/physrev.85.365} {\bibfield
  {journal} {\bibinfo  {journal} {Physical Review}\ }\textbf {\bibinfo {volume}
  {85}},\ \bibinfo {pages} {365} (\bibinfo {year} {1952})}\BibitemShut
  {NoStop}%
\end{thebibliography}%

%\section*{Author contributions}
%S.P., P.A.B., M.I.K., G.V. and A.E. conceived and designed the manuscript. S.P., P.A.B. and A.E performed first-principles calculations.
%S.P., P.A.B., M.I.K. and A.E. wrote the manuscript. All authors discussed the data and commented on the manuscript. P.A.B. and A.E. supervised the project.
%\section*{Competing interests}
%The authors declare no competing interests. 

\end{document}